\documentclass[12pt]{article}

\usepackage{amsmath}
\usepackage{graphicx}

\usepackage[letterpaper,margin=1in]{geometry}

\renewenvironment{abstract}
	{\quotation}
	{\endquotation}

\date{}

\makeatletter
\renewcommand{\fnum@figure}{\textbf{Figure \thefigure}}
\renewcommand{\fnum@table}{\textbf{Table \thetable}}
\makeatother

\usepackage{scicite}

\usepackage{url}

\let\oldhat\hat
\renewcommand{\vec}[1]{\boldsymbol{\mathbf{#1}}}
\renewcommand{\hat}[1]{\oldhat{\mathbf{#1}}}

\makeatletter
\DeclareTextCompositeCommand{\r}{OT1}{A}{%
  \leavevmode\vbox{%
    \offinterlineskip
    \ialign{\hfil##\hfil\cr\char23\cr\noalign{\kern-1.15ex}A\cr}%
  }%
}
\makeatother

\newcommand{\angstrom}{\text{\r{A}}}

\newcommand{\eAsq}{$e^{-}$/\r{A}$^2$}
\def\WSe#1{WSe\textsubscript{2}}
\def\ang#1{#1$^\circ$}
\newcommand{\degreeCelsius}{$^\circ$C}
\DeclareMathOperator{\ReLU}{ReLU}

\usepackage{changepage}

\def\scititle{
	3D Mapping of Defects and Moir\'e Corrugations via Electron Ptychography Atomic Coordinate Retrieval
}
\title{\bfseries \boldmath \scititle}

\author{
	Jeffrey~Huang$^{1,2\dagger}$,
	Yichao~Zhang$^{1,2,3\dagger}$,
	Sang~hyun~Bae$^{1,2}$,\and
    Ballal~Ahammed$^{2,4,5}$,
	Elif~Ertekin$^{2,4,5}$,
	Pinshane~Y.~Huang$^{1,2,5\ast}$\and
}

\begin{document}

\maketitle

\vspace{-0.75cm}

\begin{adjustwidth}{12pt}{12pt}
\setlength{\parindent}{0cm}
	\small$^{1}$Department of Materials Science and Engineering, University of Illinois Urbana-Champaign, IL 61801, United States.
	
	\small$^{2}$Grainger College of Engineering, University of Illinois Urbana-Champaign, IL 61801, United States.
	
    \small$^{3}$Department of Materials Science and Engineering, University of Maryland, College Park, MD~20742, United States.
    
    \small$^{4}$Department of Mechanical Science and Engineering, University of Illinois Urbana-Champaign, IL 61801, United States.
    
    \small$^{5}$Materials Research Laboratory, University of Illinois Urbana-Champaign, IL 61801, United States.
    
\begin{center}
	\small$^\ast$Corresponding author. Email: pyhuang@illinois.edu
	
	\small$^\dagger$These authors contributed equally to this work.
\end{center}
\end{adjustwidth}

\begin{abstract} \bfseries \boldmath
Defects and reconstructions in 2D moir\'e materials cause out-of-plane deformations which strongly modify their electronic properties but are difficult to experimentally access.
Here, we solve the 3D atomic coordinates of twisted bilayer \WSe2 with picometer-scale accuracy using multislice electron ptychography (MEP) acquired from a single orientation. The resulting atomic models individually visualize each of the six atomic planes, revealing the curvature of each \WSe2 layer, variations in the interlayer spacing, and the 3D locations of individual vacancies--which lie exclusively in the outer Se planes.  We also observe a new, unexpected type of structural disorder consisting of mixed bending- and breathing-type moir\'e-induced corrugations that should strongly impact the emergent electronic properties.
Broadly, our methods generate 3D atom-by-atom models of a 2D heterointerface from data acquired in about 30 seconds, methods that should unlock routine access to 3D atomic information in 2D systems and catalyze design methods to control out-of-plane deformations.
\end{abstract}

\section*{Introduction}

\noindent
Three-dimensional (3D) structural relaxations have profound effects on the properties of interfaces across materials systems. For example, in moir\'e materials such as bilayers of graphene or transition metal dichalcogenides (TMDCs) with small interlayer twists, the atoms displace in-plane to increase regions of low-energy stacking and reduce areas of the higher-energy stacking \cite{Alden_PNAS,Yoo_TBG_recon,weston_tmdc_recon,Bragg_interfero_tbl_graphene,Edelberg_Pasupathy_STM,Butz_dislocations_graphene}. These structural relaxations strongly modify the properties of 2D moir\'e systems from those of rigid layers; they have been correlated to the emergence of flat bands \cite{Naik_flat_bands_tmdc,Nuckolls_moire_review_STM} and exotic electronic phases such as superconductivity in graphene \cite{Cao_graphene_superconduct} and twisted bilayer \WSe2 \cite{Cory_WSe2_superconduct,Mak_WSe2_superconduct}. The in-plane relaxations are also disordered on length scales of 10s-100s of nm \cite{Bragg_interfero_tbl_graphene,Tillotson_tmdc_SEM}, creating spatial variations in the atomic and electronic structures that are unavoidable in moir\'e systems \cite{Ochoa_Fernandes_disorder}.

Yet, in-plane behavior does not fully capture the structural relaxations of moir\'e  systems. Modeling of 2D twisted bilayers predicts out-of-plane structural relaxations--periodic patterns of corrugation and interlayer spacing variations \cite{Maity_phason,DiXiao_ML_DFT,Maity_tbl_tmdc_recon,weston_tmdc_recon,Zhang_bilayer_theory} that accompany the in-plane displacements. Experimentally, these out-of-plane displacements are difficult to access.   Scanning probe methods such as scanning tunneling microscopy (STM) can detect corrugations only in the topmost atomic layer \cite{Brihuega_STM_tbg_corrugation,Molino_STM_antipar_WS2}. Electron tomography (ET), where 3D structures are reconstructed from a series of scanning transmission electron microscopy (STEM) images acquired from multiple tilt angles, can achieve angstrom-scale resolution in 3D--but reaches its highest 3D resolution for specific sample geometries, such as nanoparticles or needle-shaped specimens \cite{Miao_Ercius_AET_Review,Scott2012}.
For 2D materials, ET has been used to probe buckling at defects in monolayers \cite{Tian2020_scanning_AET,Tian_AET_in_plane_hetero}, but the planar geometry of 2D materials ensures the presence of missing wedge effects, limiting ET's out-of-plane resolution \cite{Ercius_ET_review_2015} and
making it difficult to probe sub-angstrom variations in interlayer spacing.

Advances in multislice electron ptychography (MEP) 
present a new route for obtaining 3D atomic structure.
Electron ptychography enables exceptionally high spatial resolutions, up to 0.2~\angstrom\ in-plane \cite{chen_ms_ptycho,Yichao_phason}, with depth resolutions of 2-4 nm \cite{chen_ms_ptycho,colum2024_3d_hbn,Ribet_MEP_NP,Zhu_LeBeau_MEP_perovskite}. While the typical nanoscale depth resolution of MEP is not sufficient for atomic resolution in 3D, standard MEP uses a completely general approach without any prior information about the sample. By augmenting 3D information from MEP with simple assumptions such as atomicity (that the sample is composed of discrete atoms) and known atomic spacings, it should be possible to extract considerably more 3D information, unlocking fast and powerful new routes to 3D atomic structures. 

\section*{Results}

\subsection*{Multislice ptychography with sub-nanometer depth resolution}

\begin{figure}[htbp!]
\centering
\includegraphics[width=0.77\textwidth]{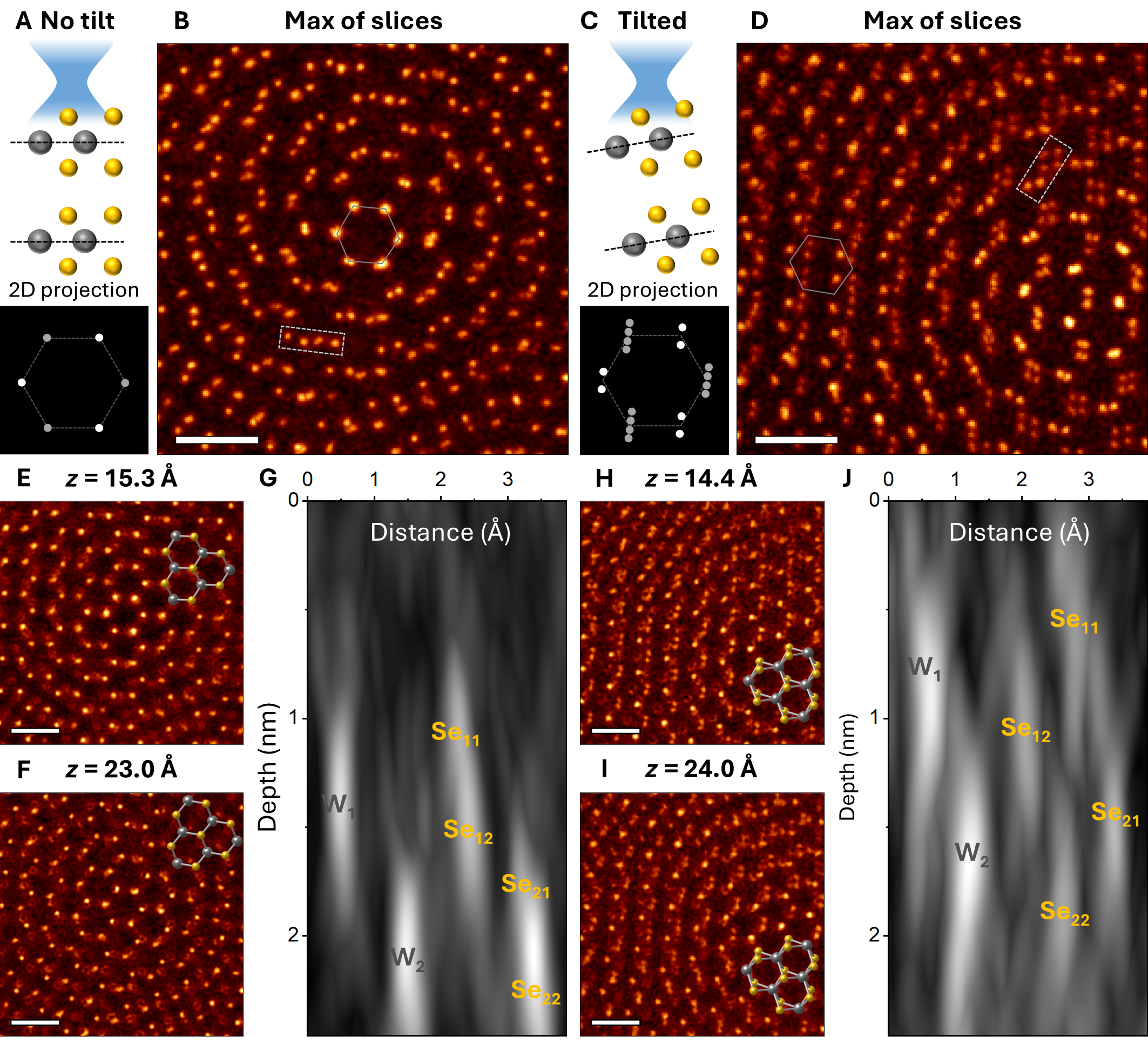}
\caption{\label{fig:DepthRes} \textbf{Distinguishing individual atoms using sub-nanometer depth-resolved multislice electron ptychography from a single orientation}. (\textbf{A} and \textbf{C}) Schematics of 2D projected atomic coordinates of a region of bilayer \WSe2 with near-AA stacking before (A) and after (C) tilting. Only 6 atomic columns are resolved in the 2D projection without tilt, while all 18 atoms are resolved in the 2D projection with tilt. (\textbf{B} and \textbf{D}) Max phase image of twisted bilayer \WSe2 from an experimental multislice electron ptychographic reconstruction without (B) and with (D) a 15-degree sample tilt. Gray hexagons mark sample regions similar to those in (A, C). (\textbf{E},\textbf{F},\textbf{H}, and \textbf{I}) Depth slices from ptychographic reconstructions without (E, F) and with (H, I) tilt, where the contrast is dominated by either the top (E, H) or the bottom (F, I) \WSe2 layer. Scale bars = 5~\angstrom. (\textbf{G} and \textbf{J}) Cross-sections of the region marked by dashed rectangles in (B) and (D), respectively. The labels indicate the atomic plane and layer number (\textit{e.g.}, Se\textsubscript{11} represents a Se atom in top atomic plane in layer 1, W\textsubscript{1} represents a W atom in layer 1). In (G), atoms in different \WSe2 layers can be distinguished by their shifts in depth but Se planes in the same \WSe2 layer are not resolved.  In contrast, in (J), all six atomic planes can be distinguished. Because the depth resolution is considerably worse than the lateral resolution, the atoms in the cross section appear elongated in the $z$ direction.}
\end{figure}

In this work, we utilize MEP augmented with priors to reveal the 3D atomic structure of twisted bilayer \WSe2. Fig.~\ref{fig:DepthRes} shows how we use MEP to distinguish individual atoms in 3D and quantify the resulting depth resolution.  We fabricate \ang{6}-twisted bilayer \WSe2 (\ang{0} refers to parallel or AB stacking) using gold-assisted monolayer exfoliation followed by bilayer assembly using a polydimethylsiloxane (PDMS) lens pickup method using the methods described in \cite{Yichao_phason} (see Supplementary Text).
Next, we collect 4D-STEM data with the sample oriented perpendicular to the electron beam (illustrated in Fig. 1A) and conduct an MEP reconstruction \cite{chen_ms_ptycho,Tsai_xray_ms_ptycho} (see Materials and Methods). In MEP, the sample is represented as a stack of thin slices, and the reconstruction solves for a 3D object whose phase represents the sample potential.

Fig. 1B shows an MEP image of \ang{6}-twisted bilayer \WSe2 (full field of view images shown in fig. S1). Specifically, Fig. 1B plots a  ``max phase'' image, a 2D projection of the 3D phase where each $(x,y)$ pixel shows the highest phase value in its depth stack. This image has an in-plane resolution of 0.31 \angstrom, as measured from the information transfer limit (see fig. S2), comparable to the highest quality reconstructions of 2D materials \cite{uncorrected_ptycho,Yichao_phason,Jiang2018}.  By leveraging the depth resolution of MEP, we separately image each \WSe2 monolayer.  Fig. 1E,F show MEP slices at different $z$ depths, where the $+z$ direction is defined along the electron beam propagation (see fig. S3 for all slices). Unlike the full 2D projection (Fig. 1B), the individual depth slices distinguish the two \WSe2 layers: Fig. 1E is dominated by the top layer, and Fig. 1F by the bottom one. 

To visualize the relative $z$ positions of the \WSe2 layers, Fig. 1G plots a cross section of the 3D phase. We estimate a depth resolution of 7.5~\angstrom\ by measuring the full width at half maximum (FWHM) in the $z$ direction for W atoms (see fig. S5). Notably, our depth resolution significantly exceeds prior experimental MEP (2-4 nm) \cite{chen_ms_ptycho,colum2024_3d_hbn,Ribet_MEP_NP,Zhu_LeBeau_MEP_perovskite}. We attribute this high depth resolution to the thinness of our sample, which enables high-quality ptychographic reconstructions from diffraction patterns with higher maximum collection angles--conditions predicted to improve MEP depth resolution \cite{zhen_high_depth_res_ptycho}. Yet, we still cannot directly resolve pairs of Se atoms, which are separated by 3.3~\angstrom\ along the crystal's $c$-axis and occupy the same lateral ($x,y$) position, so Se atom pairs appear as a single continuous column. 

To individually locate each atom, we tilt the sample by \ang{15} before acquiring 4D-STEM data (Fig. 1C). By tilting, we leverage the high lateral resolution of MEP to compensate for its lower depth resolution.  Fig. 1A,C illustrate this concept by comparing 2D projections of an AA-stacked bilayer viewed top-down (1A) versus at a tilt (1C). Whereas Se atoms are directly on top of one another when viewed top-down, tilting displaces the atoms in $x,y$. Figs. 1D,H,I show the experimental max phase image of the tilted sample (1D) and depth slices distinguishing the two \WSe2 layers (1H,I; all slices are shown in fig. S4). In the tilted view, individual Se atoms can now be distinguished. Fig. 1J shows a cross-section of the 3D phase, where each atom can be individually located and unambiguously assigned to one of the 6 atomic planes using shifts in their relative $z$ positions.

\subsection*{Retrieving 3D atomic coordinates for twisted bilayer \WSe2}

\begin{figure}[htbp!]
\centering
\includegraphics[width=0.95\textwidth]{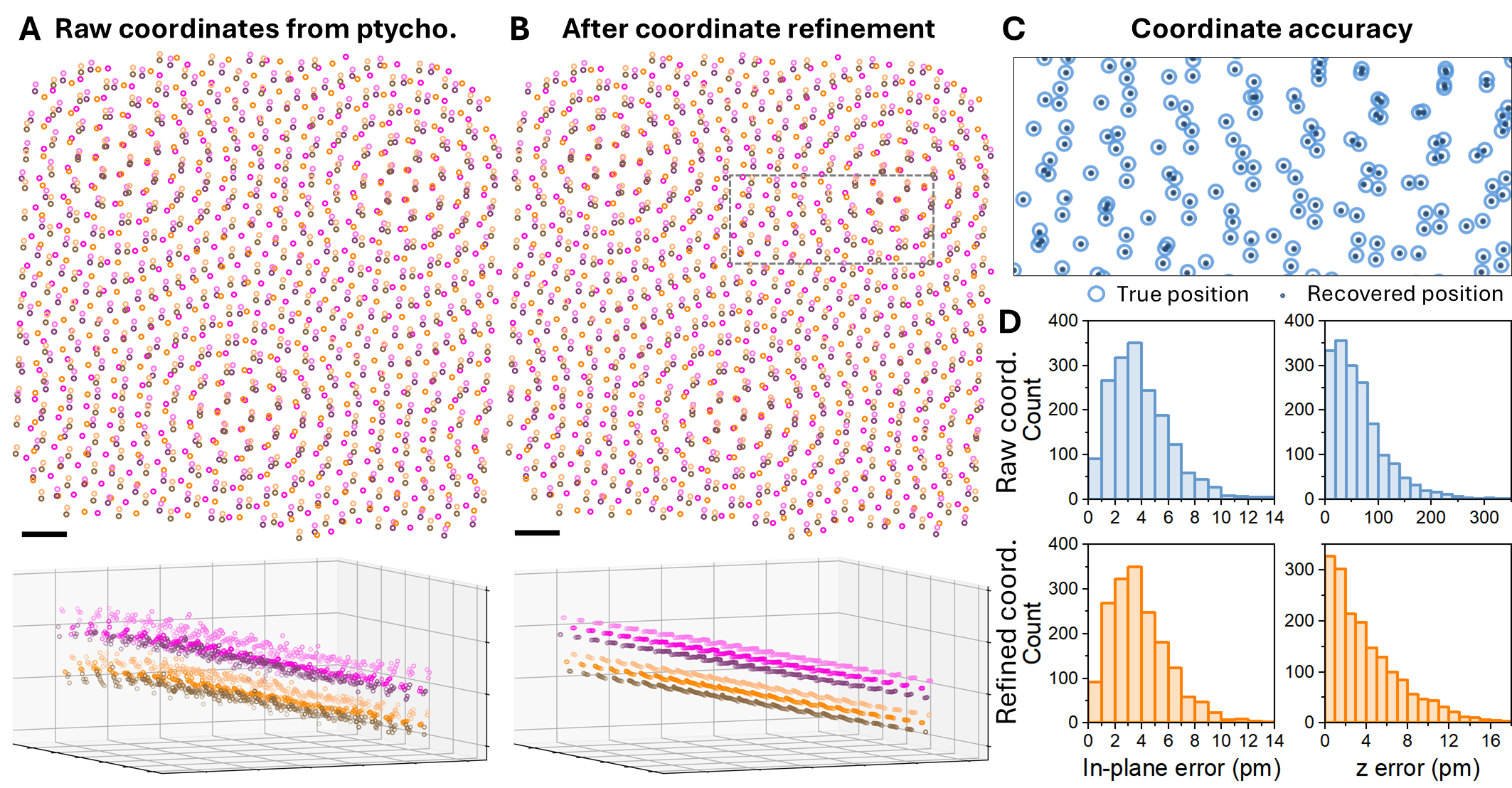}
\caption{\label{fig:Accuracy} \textbf{3D coordinate retrieval achieves pm-scale precision in all dimensions after refinement}. (\textbf{A} and \textbf{B}) Top-down view (top) and oblique view (bottom) of raw atomic coordinates taken directly from the ptychographic phase (A) and after refinement (B). Atoms are colored by the atomic layer they belong to. A corresponding ptychography image is shown in fig. S6. Scale bars = 5~\angstrom. Grid spacing in the oblique views is 1 nm. (\textbf{C}) Comparison of the true (circles, radius = 30 pm) and refined (dots) atomic positions for the region corresponding to the dashed gray box in (B). This data shows close correspondence between the true and recovered positions. (\textbf{D}) Histograms of the in-plane and $z$ errors of the 1733 refined atomic coordinates, before (top row) and after (bottom row) coordinate refinement. The $x$ and $y$ directions are perpendicular to the beam direction. In-plane error is defined as $\sqrt{dx^2 + dy^2}$, where $dx$ and $dy$ are the errors in the $x$ and $y$ directions. Before refinement, the in-plane rms error is 4.5 pm and the z rms error is 81 pm.  After refinement, the in-plane rms error is 4.5 pm and the z rms error is 5.3 pm.}
\end{figure}

Next, we develop and benchmark a workflow for 3D atomic coordinate retrieval for twisted bilayer \WSe2 from a single, tilted acquisition. We create an initial guess of the 3D atomic coordinates by finding peaks in the 3D phase. Fig. 2B shows top and oblique views of coordinates from an MEP reconstruction of a simulated 4D-STEM dataset of \ang{6}-twisted bilayer \WSe2. The 4D-STEM simulations use coordinates obtained from molecular dynamics (MD) simulations (see Materials and Methods and Supplementary Text).  By comparing retrieved coordinates against the ground truth, we see that atom-finding from MEP produces high quality $(x, y)$ atomic coordinates with a root-mean-square (rms) error of 4.5 pm (Fig. 2D, top left panel), but noisy $z$ coordinates with a rms error of 81 pm (Fig. 2D, top right panel). The larger $z$ errors are expected given our limited $z$ resolution compared to $x,y$. 

Next, we refine the 3D coordinates to be both physically plausible and consistent with our ptychography data. The refinement minimizes a loss function using the $x, y$, and to a lesser extent the $z$ coordinates from ptychography as well as prior knowledge about the structure of the sample, such as the mean W-Se bond length and average interlayer spacing (see Supplementary Text for details and uncertainty quantification). Because the $x,y$ coordinates are tightly constrained by the ptychography data, the refinement mainly affects the $z$ coordinates, which change by as much as  angstroms while the $x, y$ coordinates shift by picometers or less.

Fig. 2D (bottom left panel) shows the errors of the $x,y$ coordinates after refinement. Their 4.5 pm rms error is nearly identical to before the refinement, indicating the $x,y$ coordinates maintain high quality during refinement. Meanwhile, the refinement greatly improves the rms error of the $z$ coordinates to 5.3 pm (right column of Fig. 2D), comparable to that of the in-plane coordinates and an order of magnitude better than prior single-orientation methods \cite{Xin2010, vandyck_big_bang_tomo}. These accuracy values represent optimistic upper bounds because our 4D-STEM simulations do not include experimental factors that degrade ptychography reconstructions, such as sample drift and incoherence. Nevertheless, the simulations show that our methods can obtain 3D atomic coordinates with picometer-level accuracy using data acquired from a single view direction, demonstrating the power of augmenting MEP data with simple materials constraints.   By coupling the 3D information from MEP with our refinement process, we recover 3D atomic coordinates from a single dataset acquired in about 30 seconds, whereas ET often requires dozens of images acquired over the course of hours.

\subsection*{Mapping defects in each atomic layer}

\begin{figure}[htbp!]
\centering
\includegraphics[width=0.95\textwidth]{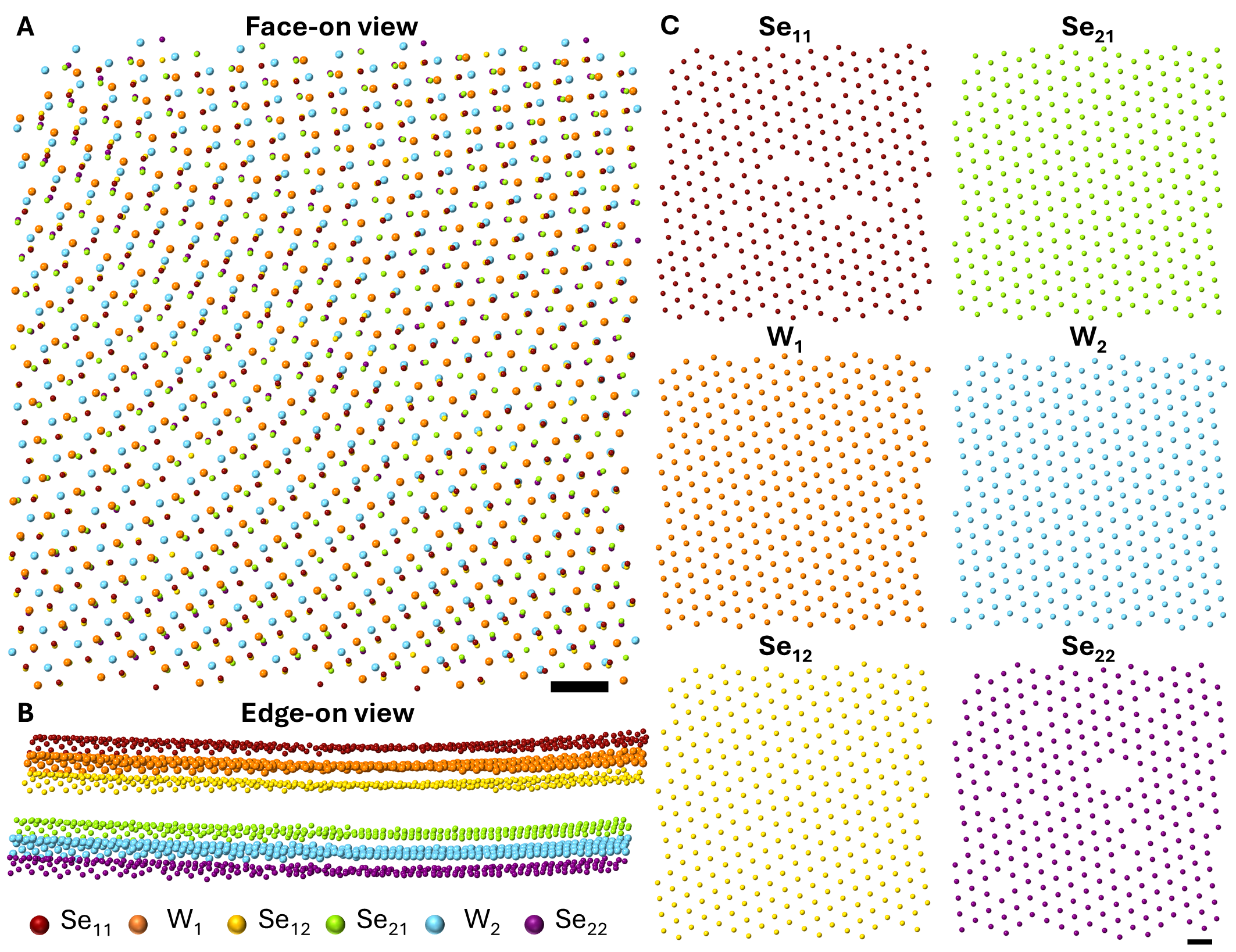}
\caption{\label{fig:Defects} \textbf{Layer-by-layer imaging and vacancy identification in \ang{2.4}-twisted bilayer \WSe2}. (\textbf{A}~and~\textbf{B}) Face-on (A) and edge-on (B) views of the atomic structure extracted from MEP, after 3D coordinate refinement. These data are acquired at a 15-degree tilt but un-tilted for the visualization. The corresponding MEP phase images are shown in fig. S7 and fig. S8. (\textbf{C}) Face-on view of atoms in each layer. Se vacancies are visible in the exterior Se$_{11}$ and Se$_{22}$ layers, but not in the interior Se$_{12}$ and Se$_{21}$ layers. These data indicate that the vacancies are not intrinsic to the bulk source \WSe2 crystals, but instead are generated during TEM sample preparation or imaging. Scale bars = 5~\angstrom\ for all panels. }
\end{figure}

In Figure 3, we extract 3D atomic coordinates from experiment and visualize individual vacancies. Figs. 3A,B show face-on and edge-on views of refined atomic coordinates from \ang{2.4}-twisted bilayer \WSe2 (corresponding ptychography reconstruction is shown in figs. S7,S8). Fig. 3C shows the atoms in each layer. Interestingly, only the outer two Se planes, labeled Se$_{11}$ and Se$_{22}$, contain vacancies. These planes contain 16 vacancies each, representing 5.0\% of the Se atoms in those layers.

Over the last decade, significant efforts have been made to understand point defects in 2D TMDCs because their type and density affect key properties such as the photoluminescence yield and charge doping, and because point defects themselves have been utilized for quantum emission, and catalysis, and memory devices \cite{Rhodes_Hone_vdW_disorder,Hus_MoS2_memristor}. (S)TEM, STM, and other scanning probe microscopies can visualize point defects in TMDCs, each with advantages and drawbacks. For example, the identities, 2D locations, and strain fields of point defects in monolayer TMDCs have measured with annular dark field STEM \cite{Chia-Hao_class_averaging_defects,Ding_WSe2_defects,Zhou_intrinsic_defect}. These studies show that chalcogen vacancies are the most prevalent point defect in synthesized TMDCs, but understanding whether chalcogen vacancies observed in (S)TEM measurements are intrinsic is challenging because electron beam damage causes vacancy formation at the doses typically used for atomic resolution imaging. Although STM and c-AFM minimize sample damage, it remains challenging to use those techniques to assign defect structures \cite{Edelberg_point_defect,Bampoulis_C-AFM_MoS2_defects,Nowakowski_c-AFM_WS2}. 

Using the 3D information in our model, we can readily identify point defects and understand their origins. Because the vacancies occur only in the outer Se layers, we believe they indicate sample damage during sample preparation or data acquisition \cite{Quincke_trilayerMoS2}, rather than intrinsic defects from the growth process. Indeed, the high number of Se vacancies is likely because we acquired three datasets in this region. The accumulated dose was 2-4 $\times 10^6$ \eAsq, typical for atomic resolution (S)TEM and sufficient to create Se vacancies \cite{Leiter_WSe2_damage}. Because the individual \WSe2 layers are exfoliated from a bulk crystal grown by chemical vapor transport, 
intrinsic defects formed during the growth of the source crystal would be expected to occur with equal probability throughout all atomic layers. These results contradict reports that the majority of Se vacancies observed in STEM represent intrinsic growth defects \cite{Ding_WSe2_defects}, while substantiating STM reports that intrinsic Se vacancies are relatively rare in CVT-grown crystals \cite{Edelberg_point_defect}.

\subsection*{Mapping out-of-plane structural reconstruction in twisted bilayer \WSe2}

\begin{figure}[bp!]
\centering
\includegraphics[width=0.83\textwidth]{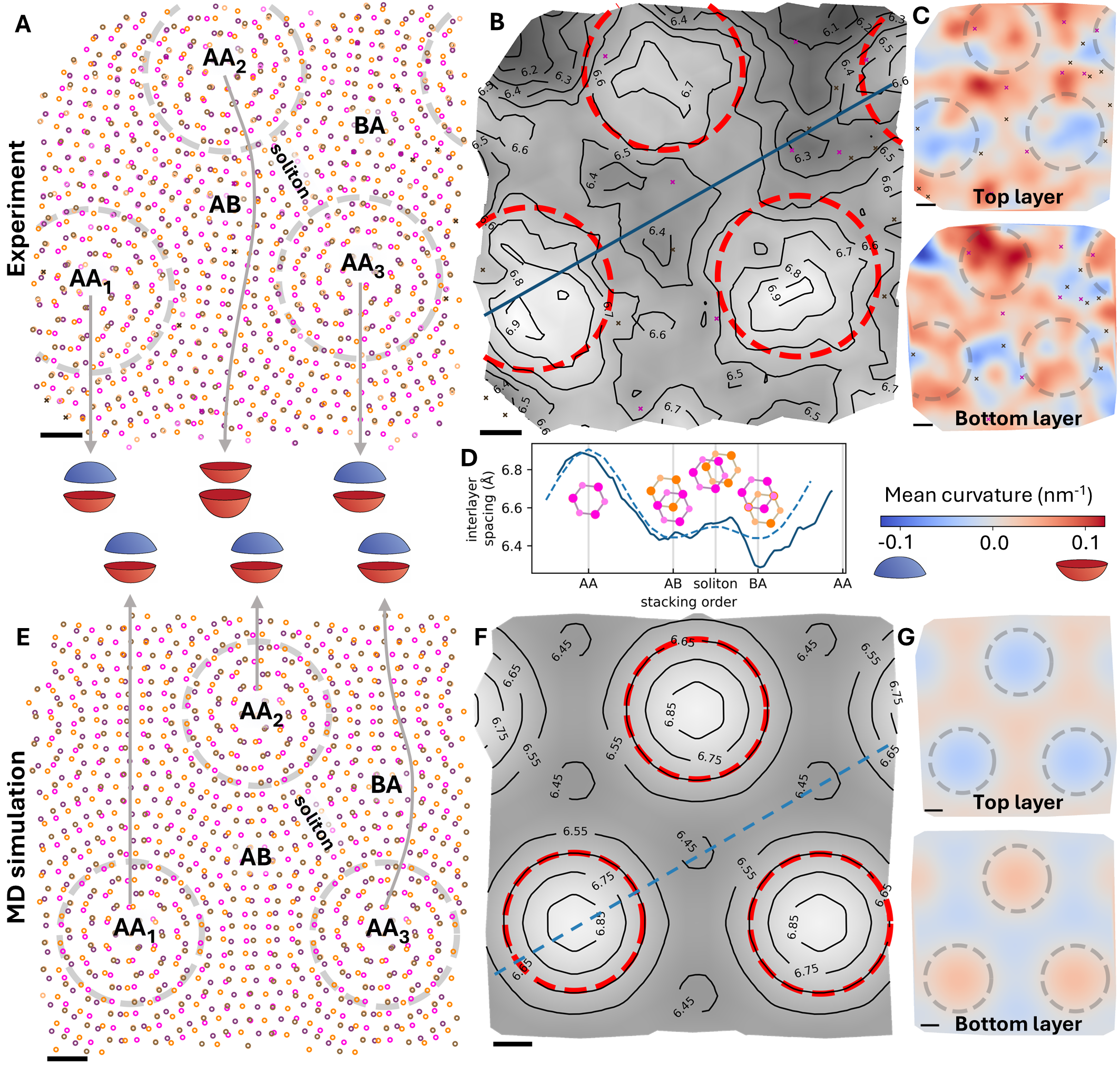}
\caption{\label{fig:Popping} \textbf{Out-of-plane deformations of twisted bilayer \WSe2}. (\textbf{A} and \textbf{E}) 2D projected atomic coordinates of \ang{6}-twisted \WSe2 obtained from experiment (A) and MD simulations (E). To determine the stacking order, all atoms are projected to a common surface defined by one of the W layers (see Materials and Methods). Locations with approximately AA stacking are labeled with dashed circles. (\textbf{B} and \textbf{F}) Contour plots of interlayer spacing between the two W layers in units of angstroms. AA stacked regions are labeled with dashed red circles. (\textbf{D}) A line profile shows the interlayer spacing is greatest in the AA-like stacking regions, lowest in the AB-like stacking regions, and slightly larger in soliton regions in both simulation (dashed line) and experiment (solid line). (\textbf{C} and \textbf{G}) Mean curvature of top (top) and bottom (bottom) W atomic layers. The curvature of each AA region is also illustrated below (A) and above (E). In the MD simulated structures, the mean curvature of AA regions takes opposite sign in the two layers, indicating that the AA regions bulge outwards in a breathing mode.  In experiment, we observe breathing modes (Regions AA$_1$ and AA$_3$) as well as bending modes (Region AA$_2$), where the two layers bend in the same direction.  Scale bars = 5~\angstrom\ for all panels.}
\end{figure}

In Figure 4, we investigate the out-of-plane structural reconstructions of \ang{6}-twisted bilayer \WSe2. Fig. 4A,E show top-down views of atomic structures obtained from  experiment (Fig. 4A) and MD simulations (Fig. 4E). In each structure, we observe three AA-like stacking regions, indicated by dashed gray circles; in these regions, atoms in one layer nearly overlie atoms of the same type in the other layer. These regions are separated by AB stacking, where the metal atoms from one layer overlap the chalcogen atoms from the other, and solitons, stacking faults that localize shear and strain from the structural reconstruction \cite{Maity_tbl_tmdc_recon,Heine_tmdc_hetero_corrugation,Alden_PNAS,Yoo_TBG_recon,weston_tmdc_recon}.
Figs. 4B,D,F show contour maps (Figs. 4B,F) and corresponding line profiles (Fig. 4D) of the interlayer spacing through different stacking orders. In experimental data, the interlayer spacings are greatest in the AA regions, reaching up to 6.7-6.9~\angstrom, 0.4-0.6~\angstrom\ larger than that of the AB regions closest to them (6.2-6.4~\angstrom). The solitons also exhibit a slightly larger (by 0.1~\angstrom) interlayer spacing compared to AB. These data are in excellent quantitative agreement with MD simulations, where the interlayer spacings reach a maximum of 6.9~\angstrom\ in the AA regions and minimum of 6.4~\angstrom\ in AB regions (Fig. 4E,F).  Similar results have been reported in prior theoretical studies \cite{weston_tmdc_recon,Maity_phason,DiXiao_ML_DFT,Maity_tbl_tmdc_recon}, but have been difficult to validate because of a lack of methods to probe local changes in interlayer spacing. These variations occur because the AA and soliton stackings are energetically unfavorable compared to the AB stacking configuration \cite{Alden_PNAS,Yoo_TBG_recon,Naik_flat_bands_tmdc,naik_kolmogorov}, resulting in local interlayer repulsion.

Finally, we probe out-of-plane corrugations in twisted bilayer \WSe2. To understand how the moir\'e structure bends to accommodate changes in interlayer spacing, we examine the sign of the mean curvature of the AA regions.  The mean curvature measures how a surface locally bends; it is the average at each point of the curvature in two principal directions (see Materials and Methods for description of this analysis and fig. S11 for maps of the principal curvatures).  We plot the mean curvature of the top and bottom W planes for experiment (Fig. 4C) and MD simulations (Fig. 4G).  In MD simulations (Fig. 4G), the mean curvature at each of the three AA regions is negative (blue), indicating domes in the top \WSe2 layer, but positive (red), indicating bowls in the bottom \WSe2 layer. This behavior corresponds to  an ``out-of-phase'' or ``breathing mode'' deformation, where the local increase in interlayer spacing at AA regions occurs by bulging the two layers outwards.  We qualitatively recover the same curvatures in ptychographic reconstructions of MD structures (fig. S12). Fig. 4C shows the experimental curvatures for the \ang{6}-twisted \WSe2.  In the regions labeled AA$_1$ and AA$_3$, both \WSe2 layers bulge outwards, similar to the MD simulations in Fig. 4E-G.  In contrast, at AA$_2$, the top and bottom layers both appear red, indicating that they are curved in the same direction, forming an ``in-phase'' or ``bending mode'' deformation.

Theoretical studies on twisted bilayer 2D materials have predicted both in-phase and out-of-phase corrugations \cite{Heine_tmdc_hetero_corrugation, Tawfiq_corrugation,Zhang_bilayer_theory,universal_moire_buckling} similar to those observed in our experiment. These predictions show that the twist angle and layer composition determine whether in- or out-of-phase modes represent the minimum-energy configuration \cite{Tawfiq_corrugation, Heine_tmdc_hetero_corrugation}. Studies of twisted TMDC homobilayers generally predict out-of-phase corrugations \cite{Heine_MoS2_corrugation,Zhang_bilayer_theory} at every AA region. In contrast, mixing of in- and out-of-phase modes in close proximity, as we observed, has neither been previously predicted nor experimentally observed. We attribute this mixing to the small energy difference between the two states. For example, for twisted bilayer graphene, this energy difference is \(\sim \)0.01 meV per atom for few-degree twists \cite{Tawfiq_corrugation}. Furthermore, strain from sample fabrication and kinetic trapping may alter the sample from its thermodynamically stable configuration. These results are important because out-of-plane reconstructions have been tied to emergent superconductivity and changes in the electronic filling and vibrational states of moir\'e systems \cite{Tawfiq_corrugation}. Importantly, the out-of-plane reconstructions are spatially non-uniform, representing a new type of disorder in moir\'e systems.

\section*{Discussion}
To conclude, we use multislice electron ptychography to retrieve 3D atomic coordinates of twisted \WSe2 bilayers, locate single vacancies in 3D, and map corrugations and sub-angstrom changes in the interlayer spacing due to moir\'e reconstructions. These methods demonstrate MEP's ability to unlock previously inaccessible out-of-plane deformations at buried interfaces.  Moreover, our methods dramatically accelerate atomic-level 3D imaging for 2D materials, reducing acquisition times from hours to seconds and potentially unlocking \textit{in situ} measurements of atomic structures and their transformations in all three dimensions. For 2D materials, our work uncovers disordered 3D corrugation modes that should be important to understanding and controlling the electronic properties of moir\'e systems.

\section*{Materials and Methods}
\subsection*{Sample preparation}
Bulk \WSe2 was purchased from HQ Graphene. Monolayers of \WSe2 were exfoliated using a modified gold-assisted exfoliation method \cite{exfol2D, tmdc_exfol}, assembled into twisted bilayers, and suspended over TEM grids with $\sim$2~{\textmu}m diameter holes. See Supplementary Text for more details.

\subsection*{Data acquisition}
We characterized the samples using a Thermo Fisher Scientific Themis Z aberration-corrected scanning transmission electron microscope. The microscope was operated at 80 kV to minimize electron beam damage. We used a defocused electron probe (7.5~nm and 10~nm overfocus) with a probe current of 7-15 pA and convergence semiangle of 25 mrad to acquire 4D-STEM data sets on a first generation EMPAD detector with $128 \times 128$ scan positions using a scan step size of 0.43~\angstrom\ and a dwell time of 1 millisecond (ms). As timed with a stopwatch, a 4D-STEM data set with $128 \times 128$ scan positions took $\sim$33.5~s to record, corresponding to 2.0 ms per diffraction pattern -- this is mostly accounted for by the 1 ms dwell time used and the 0.86 ms per frame read time \cite{tate_EMPAD}.  The camera length is 230 mm corresponding to a maximum scattering k-vector of 2.5~\angstrom$^{-1}$.  For the tilted experiments (Figures 1D, 3, and 4A), an alpha tilt of 15 degrees was applied to the sample holder.

\subsection*{Ptychography}
Each 4D-STEM dataset consists of an array of convergent beam electron diffraction (CBED) patterns. A multislice ptychographic reconstruction of the sample is produced by matching the experimental diffraction patterns to simulations using a forward model of electron scattering through a 3D sample.
We performed ptychographic reconstructions using the \textsc{fold\textunderscore slice} package \cite{chen_ms_ptycho, foldslice}, a fork of the \textsc{PtychoShelves} package \cite{PtychoShelves}.  For all reconstructions, the multislice maximum likelihood solver was used \cite{Tsai_xray_ms_ptycho, Thibault_max_likelihood}.  After every iteration of the reconstruction, the object was regularized according to the weighting function described in Ref.~\citenum{colum2024_3d_hbn}, with the layer regularization parameter $\gamma$ set to 0.1.  Diffraction patterns were padded to increase the real space sampling of the ptychographic reconstruction.  Probe position correction was enabled.  For reconstructions on experimental data, multiple probe modes were used to model partial coherence \cite{Thibault_mixed_state}.

Parameters for the final reconstructions are as follows:
\begin{itemize}
    \item Fig. 1B: Diffraction patterns padded to $256 \times 256$ pixels; 8 probe modes, 19 slices, 1.535~\angstrom\ spacing between slices.
    \item Fig. 3: Diffraction patterns padded to $192 \times 192$ pixels; 8 probe modes, 23 slices, 1.5~\angstrom\ spacing between slices.
    \item Fig. 2 and fig. S12: Diffraction patterns padded to $192 \times 192$ pixels; 1 probe mode, 22 slices, 1.6~\angstrom\ spacing between slices. This reconstruction was on simulated data.
    \item Figs. 1D and 4A: Diffraction patterns padded to $192 \times 192$ pixels; 10 probe modes, 22 slices, 1.6~\angstrom\ spacing between slices.
\end{itemize}

The max phase image of Fig. 1B was formed by taking the 7th through 19th object slices, which contained the \WSe2 sample, and taking the maximum phase value along the $z$ direction (the direction of the beam). The max phase image of Fig. 1D was formed from the 3rd through 20th object slices.

\subsection*{Coordinate retrieval}
Generally, the peaks in the phase of the 3D object function produced by multislice ptychography correspond to atoms. An initial list of peaks in the 3D phase is generated by listing voxels that are local maxima -- that is, voxels that have the greatest phase value in their local $3 \times 3 \times 3$-voxel neighborhood.  Next, a threshold is applied to remove dim voxels that are only background noise.  Then, for each of these peaks, a more precise peak position with subpixel precision is found by looking for a local maximum in the tricubic interpolation of the phase (see Supplementary Text) around the voxel.  To do this, the tricubic interpolation of the phase is represented as a piecewise function of 3D position, and our program compares the interpolated phase values of all nearby points where the gradient of this function is zero. The point with the highest phase is deemed to be the more precise peak position.

This approach works for locating most isolated atoms, but fails for atoms that are overlapping in the 3D phase produced by ptychography, as well as for atoms that are close to the top or bottom face of the reconstruction volume.  For such atoms, we resort to estimating atom positions by eye.  We also manually add the locations of the lattice site wherever there is a vacancy.

Next, we label each position by the atomic layer that it belongs to.  This is made possible by the tilted sample orientation and depth resolution of the multislice reconstruction.  Finally, the coordinates are refined as described in the Supplementary Text.

\subsection*{Analysis of coordinates}
We constructed for each W layer a triangular mesh from the W coordinates using the \textsc{Open3D} package \cite{Zhou_Open3D}. In Figs. 4A,E, for the sake of illustrating the actual stacking order, each atom is projected to its nearest point within the mesh surface passing through the higher W layer. Rigidly rotating the atomic model is not enough to accurately show the stacking order for Fig. 4A, because the \WSe2 layers are curved.

\subsubsection*{Interlayer spacing}
To calculate interlayer spacing as a function of position, we sample a grid of points on the triangular mesh of the higher W layer and compute the closest distance between these points and the lower W mesh.  Mean interlayer spacing is calculated as the average over this grid of points.

\subsubsection*{Estimate of curvature}
For a point on a surface, there exist two principal directions where curvature is at a maximum or minimum, and for each principal direction, there is a principal curvature value defined as the reciprocal of the radius of the osculating circle passing through the point along that direction. Starting from the triangular meshes of each W layer, we use the method described in \cite{estimate_curv} to calculate the normal vector, principal curvature values, and principal directions at each W atom.
The mean curvature values (the average of the two principal curvature values) are interpolated to produce the plots in Fig. 4C,G and fig. S12C,D.

\subsection*{Molecular dynamics simulations}
We perform classical MD simulations via LAMMPS \cite{LAMMPS_2022} using a hexagonal unit cell of 6-degree twisted bilayer \WSe2. We use the Stillinger-Weber potential to describe the intralayer interactions and Kolmogorov-Crespi potential \cite{naik_kolmogorov} for interlayer interactions.  We use the parameters for the aforementioned force fields from \cite{naik_kolmogorov}. We first equilibrate the structure at 300 K for 100 ps with a Nose-Hoover thermostat using a time step of 0.5 femtoseconds. We apply a tether to the center of mass of the system to resolve the flying ice cube problem \cite{flying_ice_cube}. After equilibration, the system is relaxed for another 100 ps to transition into microcanonical ensemble. The simulation continues for 3 nanoseconds, and we calculate an average atomic position.


\clearpage 

\providecommand{\noopsort}[1]{}\providecommand{\singleletter}[1]{#1}%


\section*{Acknowledgments}
We thank D. Xiao, T. Cao, and K. Cauffiel for helpful discussions.

\paragraph*{Funding:}
This work was supported by the Air Force Office of Scientific Research under award number FA9550-20-1-0302 and the National Science Foundation under award number DMR-2309037. J.H. is partially supported by the DIGI-MAT program through National Science Foundation under Grant No. 1922758. Electron microscopy facilities were provided by the University of Illinois Materials Research Laboratory. We acknowledge use of microscopy facilities funded by an NSF-MRSEC under award number DMR-2309037. This work made use of the Illinois Campus Cluster, a computing resource that is operated by the Illinois Campus Cluster Program (ICCP) in conjunction with the National Center for Supercomputing Applications (NCSA) and which is supported by funds from the University of Illinois Urbana-Champaign. This work also used the Delta advanced computing and data resource at the NCSA at University of Illinois Urbana-Champaign through allocation MAT240032 from the Advanced Cyberinfrastructure Coordination Ecosystem: Services \& Support (ACCESS) program, 
which is supported by National Science Foundation grants \#2138259, \#2138286, \#2138307, \#2137603, and \#2138296.

\paragraph*{Author contributions:}
Y.Z. acquired all 4D-STEM data. J.H. performed reconstruction of all ptychography images and subsequent image analysis. S.H.B. prepared the twisted bilayer samples and transfer to TEM grids. B.A. conducted MD simulations, supervised by E.E. P.Y.H supervised the project. J.H., Y.Z., and P.Y.H. wrote the paper. All authors edited the paper.

\paragraph*{Competing interests:}
There are no competing interests to declare.

\paragraph*{Data and materials availability:}
All data will be made available upon publication.


\subsection*{Supplementary materials}
Supplementary Text\\
Figs. S1 to S12\\
References \textit{(62-\arabic{enumiv})}\\ 


\newpage


\renewcommand{\thefigure}{S\arabic{figure}}
\renewcommand{\thetable}{S\arabic{table}}
\renewcommand{\theequation}{S\arabic{equation}}
\renewcommand{\thepage}{S\arabic{page}}
\setcounter{figure}{0}
\setcounter{table}{0}
\setcounter{equation}{0}
\setcounter{page}{1} 


\begin{center}
\section*{Supplementary Materials for\\ \scititle}

Jeffrey~Huang$^{\dagger}$,
Yichao~Zhang$^{\dagger}$,
Sang~hyun~Bae,\\
Ballal~Ahammed,
Elif~Ertekin,
Pinshane~Y.~Huang$^{\ast}$\\ 
\small$^\ast$Corresponding author. Email: pyhuang@illinois.edu\\
\small$^\dagger$These authors contributed equally to this work.
\end{center}

\subsubsection*{This PDF file includes:}
Supplementary Text\\
Figures S1 to S12\\


\newpage


\subsection*{Supplementary Text}

\subsubsection*{Sample preparation}

\paragraph*{Exfoliation of \WSe2 onto silicon dioxide wafer:}
For the samples in Figures 1B and 3, the 100 nm gold film was deposited onto a flat silicon wafer. A thin layer of poly(methyl methacrylate) (PMMA) (495k 2\% in anisole purchased from MicroChem Corp.) was spun on the now gold-coated wafer at 3000 rpm for 1 minute. The wafer with gold and PMMA was cured at 150~\degreeCelsius\ for 5 minutes. 0.5~cm $\times$ 0.5~cm thermal release tape (TRT) with a release temperature of 90~\degreeCelsius\ was adhered atop the coated PMMA as a handling layer. Flat bulk \WSe2 was affixed to a flexible plastic surface using double-sided Kapton tape. The bulk crystal was cleaved using 3M scotch tape to expose a clean surface. The gold / PMMA / TRT construct was peeled off the wafer, quickly placed on the affixed bulk \WSe2, and then firmly pressed. The construct was carefully peeled off, with the top layer of the bulk crystal adhered to the gold. The construct was placed on a silicon / silicon dioxide wafer with 285~\angstrom\ oxide thickness, firmly pressed, then placed in a vacuum chamber to remove bubbles and improve adhesion. The wafer was placed on a hot plate at 100~\degreeCelsius\ to release TRT. Once the TRT was released, the wafer was placed in three sequential acetone baths for 1 hour each, two isopropyl alcohol baths for 15 minutes each, and then plasma-cleaned. The gold layer was etched with potassium iodide solution. The wafer was then placed in an active flow DI water bath for 20 minutes.

For the sample in Figure 4A, a layered 100~nm gold and 500~nm copper film were deposited on a flat silicon wafer. The film was lifted off the wafer with double-sided Kapton tape, where only one side of adhesive was exposed and a square window was cut in the middle of the tape. The Kapton tape acted as a handling layer. This gold-copper film with the Kapton handling layer was then brought into contact with bulk \WSe2, with the gold side towards the crystal. The \WSe2 crystal, with the metal film and the Kapton tape, was heated to 160~\degreeCelsius, to ensure good adhesion. The metal film with Kapton tape handling layer was then carefully lifted off the bulk crystal, now with a monolayer of \WSe2 exfoliated and adhered to the film. The film was placed on silicon / silicon dioxide wafer with 285~\angstrom\ oxide thickness, where the metal film is carefully cut along the edges off the handling layer of Kapton tape. The wafer was then placed in a vacuum chamber to remove bubbles. The copper and the gold were subsequently etched using ammonium persulfate and potassium iodide solution, each etch followed by an active flow DI water bath for 20 minutes.

\paragraph*{Construction of suspended \WSe2 twisted bilayer heterostructure on TEM grid:}
For the samples in Figures 1B and 3, twisted bilayer heterostructure was constructed by repeating the exfoliation process with gold / PMMA / TRT construct, from the same source crystal, twice onto the same wafer. The TRT handling layers were cut identically. The two exfoliations were done so that the edge of both TRT layers were aligned on the same edge of the wafer.  Once the twisted bilayer was constructed, a thin layer of 495k 2\% PMMA was spun on top of the wafer at 3000 rpm for 1 minute, then dried overnight. The PMMA layer outside the borders of the heterostructure was scraped off with a razor blade, exposing the oxide layer of the wafer. A droplet of aqueous 5M potassium hydroxide solution was pipetted onto the exposed part of the wafer to etch the oxide layer and separate the twisted bilayer and the support PMMA film from the wafer. The PMMA / twisted bilayer film was carefully floated on a beaker full of DI water to dilute the potassium hydroxide solution. The PMMA side of the floating film was brought into contact with a planar PDMS lens affixed to the edge of a glass slide which adhered the film onto the PDMS. The PDMS / PMMA / twisted bilayer stamp was air dried, then immersed in several DI water baths, then air dried again. The PDMS stamp was lowered gently onto a silicon nitride TEM grid on a transfer station.  The TEM grid was heated to 120~\degreeCelsius\ for the PMMA to soften, at which point the PDMS stamp was gently raised and removed. The silicon nitride grid, now with the twisted bilayer and the PMMA residue, was placed in a series of acetone and isopropyl alcohol baths to remove polymers.

For the sample in Figure 4A, twisted bilayer heterostructure was constructed using a modified polycarbonate / polydimethylsiloxane (PC/PDMS) lens pickup method \cite{vdw_rotational_align}. A large, clean flake of monolayer \WSe2 was identified. To control the twist angle, half of the flake was then brought into contact with the PC/PDMS lens, then lifted off the wafer, ripping the flake in half. The stage was rotated to the desired twist angle, after which the lifted \WSe2 flake was brought into contact with the remaining flake on the wafer.  Both layers were lifted off the wafer and placed onto a holey TEM grid with silicon nitride support, then heated to melt the PC sacrificial film, separating the heterostructure from the PC/PDMS lens. The heated TEM grid was then placed in multiple chloroform baths to remove PC. The TEM grid underwent one additional isopropyl alcohol bath to remove carbonaceous residue, then air dried.

\subsubsection*{Interpolation of ptychography object}
We interpolate the phase using a tricubic interpolation scheme as derived in \cite{tricubic_interp} and implemented in \cite{arbinterp}. Depth resolution was quantified using line profiles through the interpolated object phase.

The cross sections shown in Fig. 1G,J use the interpolated phase. First, a series of 2D cross-sectional slices of the interpolated phase are computed within the rectangles marked in Fig. 1B,D.  These slices run parallel to each other in the direction of the long side of the rectangle.  Then, a 2D projection of these cross-sectional slices is created by taking the maximum value across the width of the rectangle, resulting in the final cross section.  Such a 2D projection allows for a single image to include a set of atoms whose $(x,y)$ positions are not collinear.

\subsubsection*{Coordinate refinement}
The coordinates were refined by minimizing this loss function:
\begin{equation}
    \begin{aligned}
    L(\vec{r}_i) &= W_{\rm LD} \sum_i W_{d,i} \left( (x_i - x_{iU})^2 + (y_i - y_{iU})^2 \right) \\
    &+ W_{\rm B} \sum_k \left( \left\| \vec{r}_{p_k} - \vec{r}_{q_k} \right\| - B \right)^2 \\
    &+ W_{\rm VS} \sum_{j\text{ near } A} W_{v,j} \left( z_j - z_{jU} \right)^2 \\
    &+ W_{\rm VO} \sum_{L \in \{11,21 \}} \sum_s \ReLU \left( s_L \left( z_{L,s} - z_{{\rm ref},L,s} \right) \right).
    \end{aligned}
\end{equation}

The parameters to be refined ($\vec{r}_i$) are the 3D coordinates of all atoms except one W atom, which is fixed to prevent the entire set of coordinates from drifting in the $z$ direction. The fixed W atom's coordinates are still present in some of the loss function terms, however. The original coordinates ($\vec{r}_{iU}$) are produced from the 3D ptychographic phase (see previous section) and also serve as the starting point of the optimization.  Vacancy positions are included in the refinement. Some atoms and vacancies near the edge of the field of view are excluded from the refinement due to having relatively poor coordinates.

Here is an explanation of all loss function terms, following their order in the loss function definition:
\begin{itemize}
    \item Lateral displacement loss: Try to preserve the original $x, y$ coordinates ($x_{iU}, y_{iU}$), which should be relatively accurate.
    \item Bond length loss: Try to preserve the W-Se bond length, $B =$ 2.5379~\angstrom. Let the 6 atomic layers of twisted bilayer \WSe2 be labeled in order of increasing $z$ coordinate as Se$_{11}$, W$_1$, Se$_{12}$, Se$_{21}$, W$_2$, Se$_{22}$.  Then, W atoms from layer W$_1$ are bonded to Se atoms in layers Se$_{11}$ and Se$_{12}$, and similarly for the other \WSe2 layer. The index $k$ ranges over all pairs of W and Se atoms that are bonded together.
    \item Vertical spacing loss: Try to preserve the original $z$ coordinates of some W atoms near the fixed W atom.
    \item Vertical ordering loss: It can be difficult for the bond length constraint to distinguish between the Se layer directly above and below a W layer. Therefore, sometimes the refinement process may put a Se atom in the wrong layer.  For example, a Se atom from the Se$_{12}$ layer may be placed in the Se$_{11}$ layer.  To mitigate this pathology of the refinement process, this term tries to penalize Se atoms that end up in the wrong layer.  In this term, $L,s$ is an index that ranges over all Se atoms from layer $L$, $\ReLU(x) = \begin{cases}
        x, x>0 \\ 0, x \leq 0
    \end{cases}$ is the rectified linear unit, $z_{{\rm ref},L,s}$ is a reference $z$ position calculated from a weighted average of nearby Se atoms from the ``opposite'' Se layer within the same \WSe2 monolayer, and $s_L$ is either $+1$ or $-1$ depending on whether the  layer $L$ of Se atom $s$ is supposed to be above or below the opposite, reference layer.
\end{itemize}

The 3D points to be refined are manually labeled as part of one of the following categories:
\begin{itemize}
    \item Good: Good quality coordinates
    \item Poor: Poor quality coordinates, whether due to multiple atoms overlapping, or due to artifacts in the ptychographic reconstruction.
    \item Vacancy: This is a vacancy, and therefore its initial position was an estimate and not the position of a peak in the phase.
\end{itemize}
This labeling is shown for the Figure 2 data set in fig. S9.

We consider the W-Se bond length and the $x, y$ coordinates from ptychography to be known accurately, and the $z$ coordinates from ptychography to be relatively inaccurate, and weight the terms of the loss function accordingly.

The coordinates $\vec{r}_i$ are in units of angstroms. The constants in the loss function are defined as follows:
\begin{itemize}
    \item $W_{\rm LD} = \left(0.7 \cdot \Delta x \right)^{-2},$ where $\Delta x$ is the real space sampling of the ptychographic reconstruction in the $x$ and $y$ directions. For the coordinates in Figs. 3 and 4A, $\Delta x =$ 0.131~\angstrom, $W_{\rm LD} = $ 120.~\angstrom$^{-2}$.
    \item $W_{d,i} = \begin{cases}
        1,    & \text{if point }i\text{ is Good} \\
        0.25, & \text{if point }i\text{ is Poor} \\
        0.01, & \text{if point }i\text{ is Vacancy.} \\
    \end{cases}$
    \item $W_{\rm B} = \left( 0.14\,\angstrom \right)^{-2} = 51.0\,\angstrom^{-2}.$
    \item $B = 2.5379\,\angstrom$.
    \item $W_{\rm VS} = \left(1.65 \cdot \Delta z \right)^{-2},$ where $\Delta z$ is the sampling of the ptychographic reconstruction in the $z$ direction. For the coordinates in Figs. 3 and 4A, $\Delta z = 1.6\,\angstrom$, $W_{\rm VS} = 0.14\,\angstrom^{-2}.$
    \item $W_{v, j} = \begin{cases}
        1,    & \rho \leq d_1 \\
        \frac{d_2 - \rho}{d_2 - d_1}, & d_1 < \rho \leq d_2 \\
        0, & \text{otherwise}
        \end{cases}$.
        
        Here, $\rho = \sqrt{(x_{AU} - x_{jU})^2 + (y_{AU} - y_{jU})^2}$, $d_1 = 2.0\,\angstrom$, $d_2 = 4.5\,\angstrom$.
    \item $W_{\rm VO} = 50\,\angstrom^{-1}$.
\end{itemize}

Coordinate refinement was performed using PyTorch with the Adam optimizer \cite{adam_optim}. Initially, the optimizer is run for 20,000 iterations with a learning rate of $10^{-3}$.  Despite the presence of the vertical ordering term in the loss function, some Se atoms get placed into the wrong layer.  When there are still misplaced atoms after the optimizer has finished running, those atoms are individually shifted along the $z$ direction towards the correct layer, and the optimization is continued.  This process is repeated until all atoms are in the correct layer, the vertical spacing loss term is zero, and the loss has stopped improving under a learning rate of $10^{-4}$.

The coordinate refinement process thus far is quite accurate at finding the $z$ coordinates within one layer of \WSe2, but less accurate at determining the $z$ distance between the two layers. After coordinate refinement on the ptychography simulation (Fig. 2), coordinates of the top and bottom \WSe2 layers were separately aligned against the true coordinates, and the distributions of their respective $z$ coordinate errors are shown in fig. S10B.  The root-mean-square errors of the $z$ coordinates of the top and bottom \WSe2 layers are 4.1 and 6.1 pm respectively. A distribution of the error in mean interlayer $z$ spacing is shown in fig. S10A.  To generate this distribution, we performed the coordinate refinement process many times, each time using a different W atom as the fixed atom.  Across those separate refinements, the shape of each individual \WSe2 layer remained consistent, varying by only picometers, but there was considerable spread in the interlayer spacing along the $z$ direction, with the 10th percentile spacing being 0.16~\angstrom\ too small and the 90th percentile being 0.77~\angstrom\ too large. Due to the limited accuracy in retrieving the mean interlayer $z$ spacing through ptychography alone, for the atomic models in Fig. 2 and Fig. 4A, we ran the coordinate refinement multiple times and used the sets of refined coordinates that produced the average interlayer spacing closest to the value from the crystal structure of bulk 2H-\WSe2 (6.49~\angstrom) \cite{Mentzen_WSe2_structure}.
Here, ``interlayer $z$ spacing'' refers to the spacing along the beam direction, but ``interlayer spacing'' is measured perpendicular to the \WSe2 layers, the calculation of which is described in Materials and Methods.  In the end, the coordinate retrieval process made use of two pieces of prior information about twisted bilayer \WSe2: the average W-Se bond length, and the interlayer spacing.

\subsubsection*{Ptychography simulation}
For Fig. 2 and fig. S12, we simulated a 4D-STEM data set of twisted bilayer \WSe2 with the multislice method using the software package \textsc{Prismatic} \cite{prismatic2}.  For this simulation, we used the atomic coordinates calculated from molecular dynamics simulations, tilted by 12.74 degrees, and repeated periodically to cover a simulation cell of length 100~\angstrom\ in the $x$ and $y$ directions.  The multislice simulation used a real space pixel size of 0.03255~\angstrom\ and a slice thickness of 0.2~\angstrom. To model atomic vibrations, 20 frozen phonon configurations were used.  The ``3D potential'' option was enabled, allowing for an atom's potential to be split over multiple slices.  The 4D-STEM data set was simulated with accelerating voltage of 80 kV, convergence semiangle of 24.92 mrad, a probe with 75.0~\angstrom\ overfocus, $128 \times 128$ scan positions and a scan step size of 0.429~\angstrom.  Diffraction patterns were cropped and binned down to $128 \times 128$ pixels with a final pixel size of 0.01~\angstrom$^{-1}$. Poisson noise was then applied to the diffraction patterns to simulate an electron dose of $2.4\times 10^5$ \eAsq.  We performed ptychography on this data set, and then extracted and refined the coordinates.

\paragraph*{Rescaling $x, y$ coordinates:} Before the raw and refined coordinates are compared against the true coordinates, the $x, y$ coordinates of the coordinates are scaled by a factor of $0.991$. This rescaling was necessary to align the raw and refined coordinates well against the true coordinates.  The atomic position errors shown in Fig. 2 use the coordinates after this scaling. This slight miscalibration of the $x, y$ coordinates from ptychography appeared when the probe position correction was enabled. (The extracted coordinates from the experiment in Fig. 4A do not have the $x, y$ coordinates rescaled.)



\begin{figure}[hp!]\centering
    \includegraphics[width=.6\textwidth]{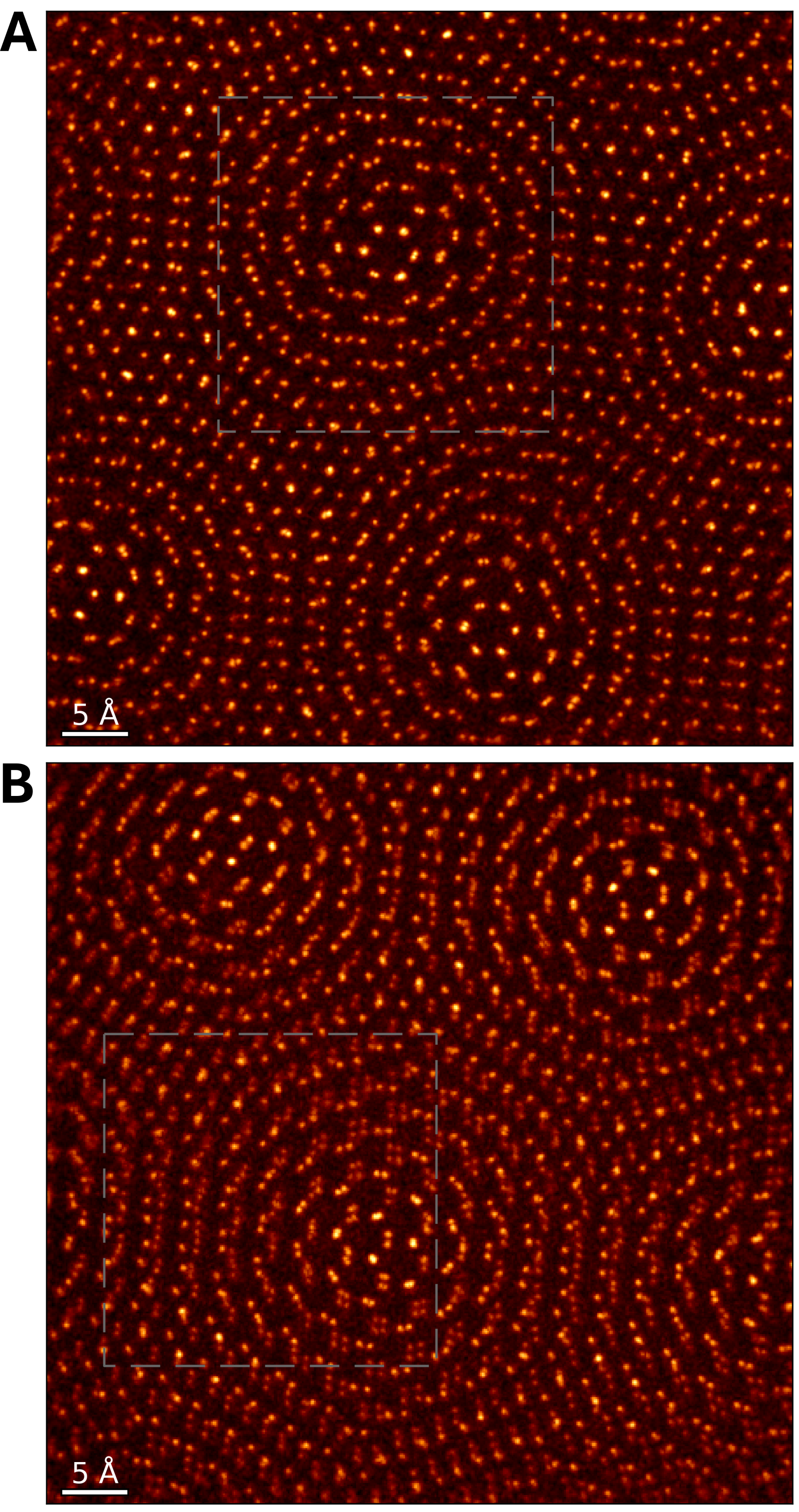}
    \caption{\textbf{Full field of view images.} (\textbf{A} and \textbf{B}) Max phase image from ptychographic reconstructions of twisted bilayer \WSe2 without (A) and with (B) sample tilt.  These are the full versions of the ptychography images presented in Fig. 1B,D, and the regions used for the cropped images are indicated by dashed rectangles. Individual Se atoms in Se columns are overlapped without sample tilt (A) but can be distinguished with sufficient sample tilt (B).}
    \label{fig:maxl_expt_full_fov}
\end{figure}

\begin{figure}[hp!]\centering
    \includegraphics[width=.95\textwidth]{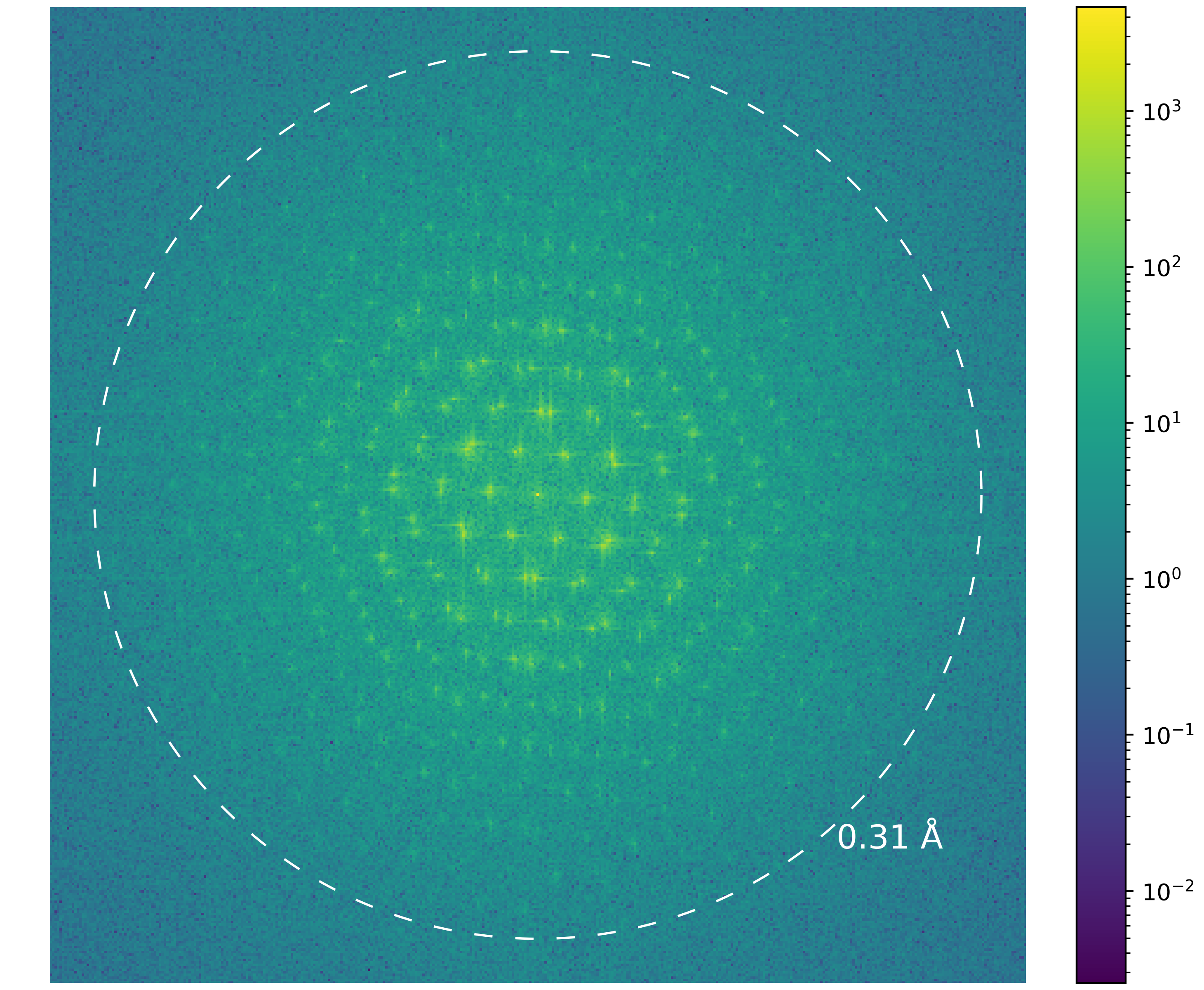}
    \caption{\textbf{Information transfer.} Cropped Fourier transform of fig. S1A (which is the full version of Fig. 1B), showing the absolute value on a logarithmic scale. The lateral spatial resolution is preserved in the max phase image.}
    \label{fig:info_transfer}
\end{figure}

\begin{figure}[hp!]\centering
    \includegraphics[width=.95\textwidth]{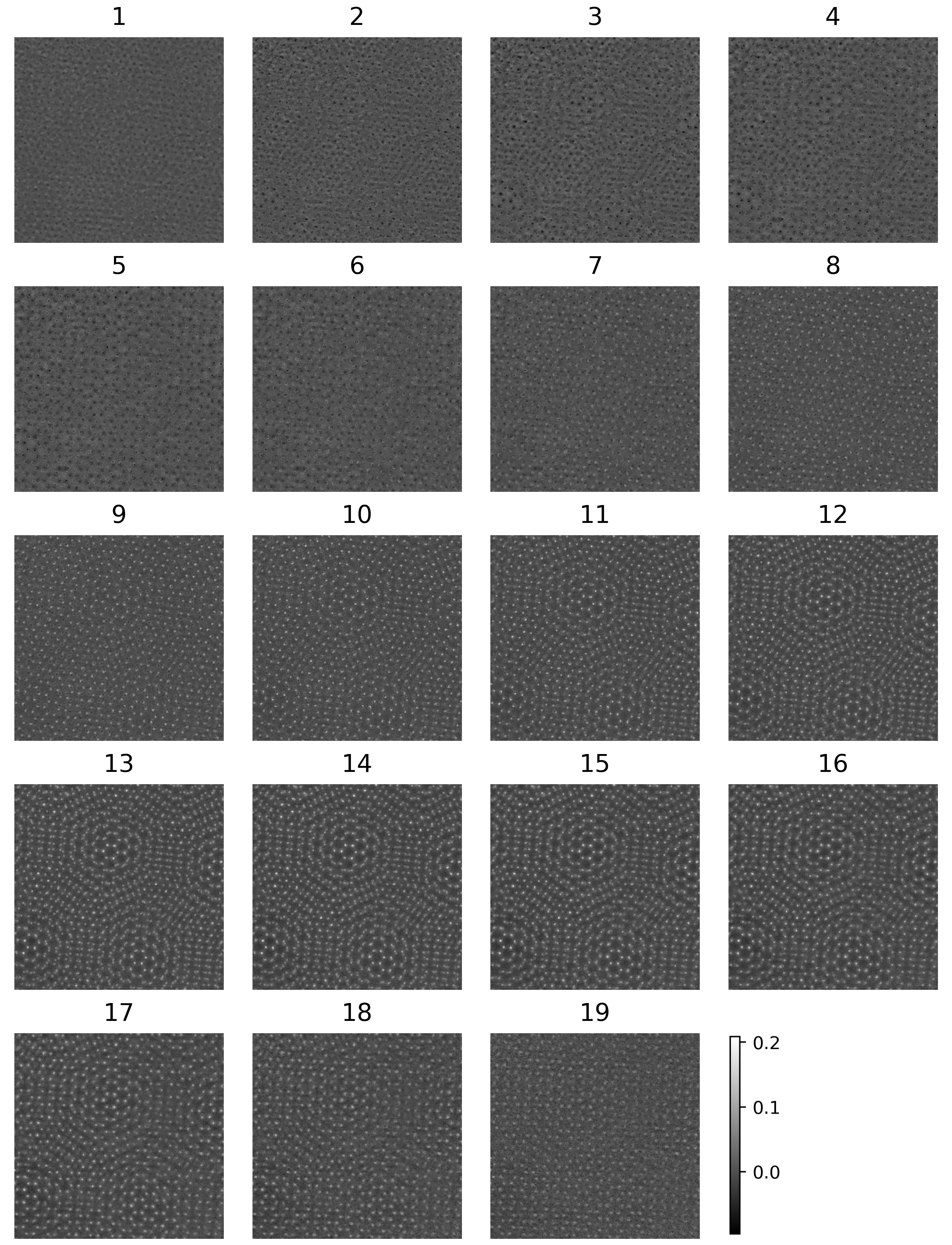}
    \caption{\textbf{Multislice reconstruction without sample tilt.} All slices of the object phase of the ptychographic reconstruction shown in part in Fig. 1B,E-G. The image is 5.53 nm wide and 5.44 nm tall. Phase values are in units of radians.}
    \label{fig:untilted_all_slices}
\end{figure}

\begin{figure}[hp!]\centering
    \includegraphics[width=.98\textwidth]{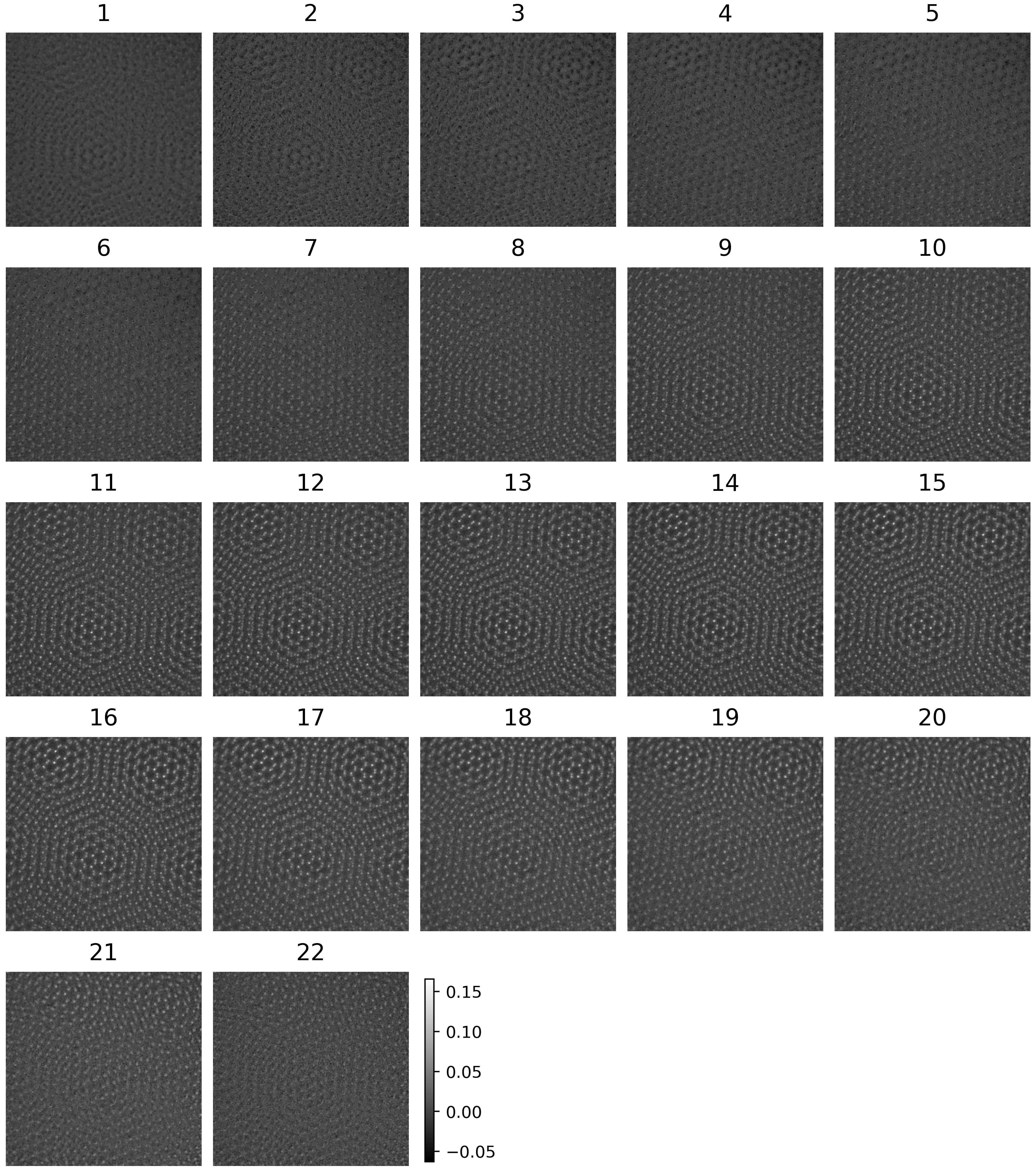}
    \caption{\textbf{Multislice reconstruction with sample tilt.} All slices of the object phase of the ptychographic reconstruction shown in part in Fig. 1D,H-J. The image is 5.58 nm wide and 5.54 nm tall. Phase values are in units of radians.}
    \label{fig:tilted_all_slices}
\end{figure}

\begin{figure}[hp!]\centering
    \includegraphics[width=.7\textwidth]{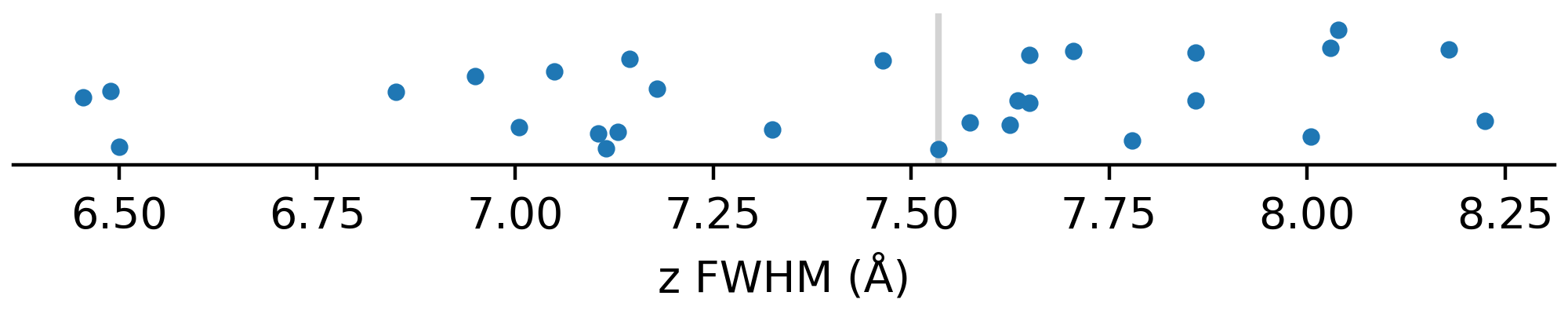}
    \caption{\textbf{Measuring depth resolution.} Dot plot of $z$-direction full width at half maximum (FWHM) values of 29 isolated W atoms from the ptychographic reconstruction shown in fig. S1A. The vertical gray line indicates the median value of 7.53~\angstrom\ reported as our depth resolution.}
    \label{fig:fwhms_z}
\end{figure}

\begin{figure}[hp!]\centering
    \includegraphics[width=.6\textwidth]{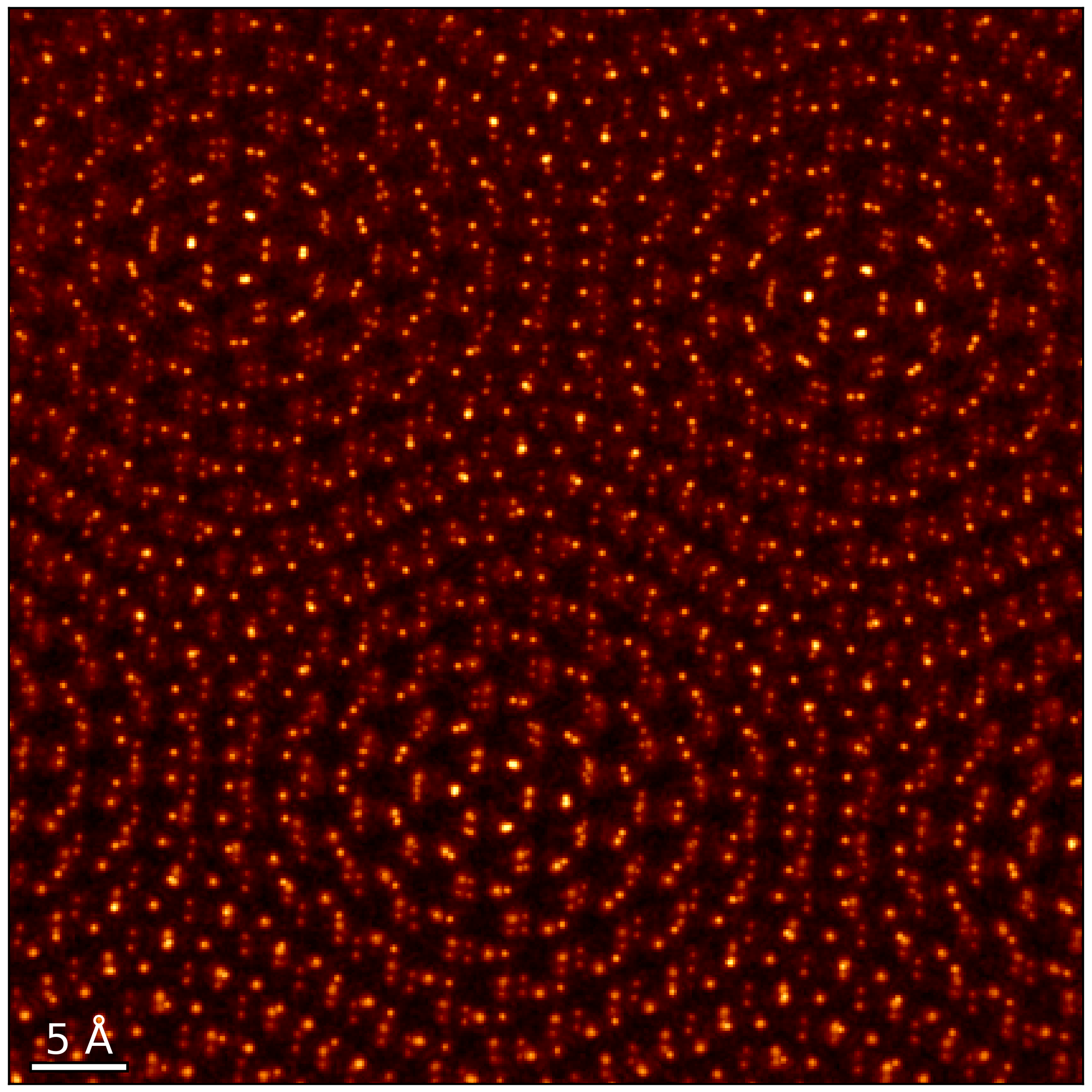}
    \caption{\textbf{Ptychography simulation.} Max phase image from the ptychographic reconstruction of twisted bilayer \WSe2 from simulated 4D-STEM data. Atomic coordinates taken from this reconstruction are shown in Fig. 2A.}
    \label{fig:ptycho_sim_maxsl}
\end{figure}

\begin{figure}[hp!]\centering
    \includegraphics[width=.6\textwidth]{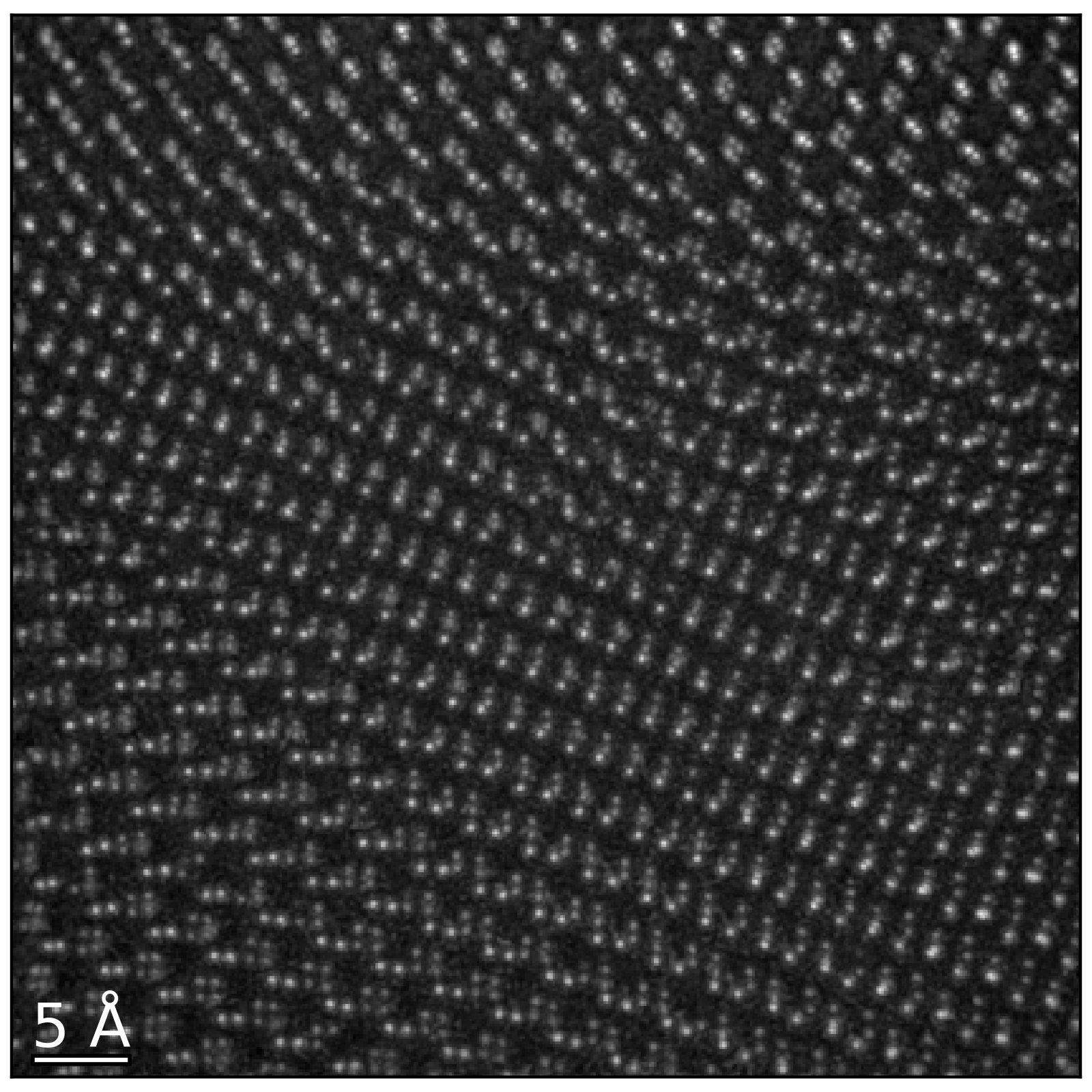}
    \caption{\textbf{Twisted bilayer \WSe2 with defects.} Max phase image of Fig. 3 from a ptychographic reconstruction of twisted bilayer \WSe2 with several Se vacancies.}
    \label{fig:maxsl_defects}
\end{figure}

\begin{figure}[hp!]\centering
    \includegraphics[width=.98\textwidth]{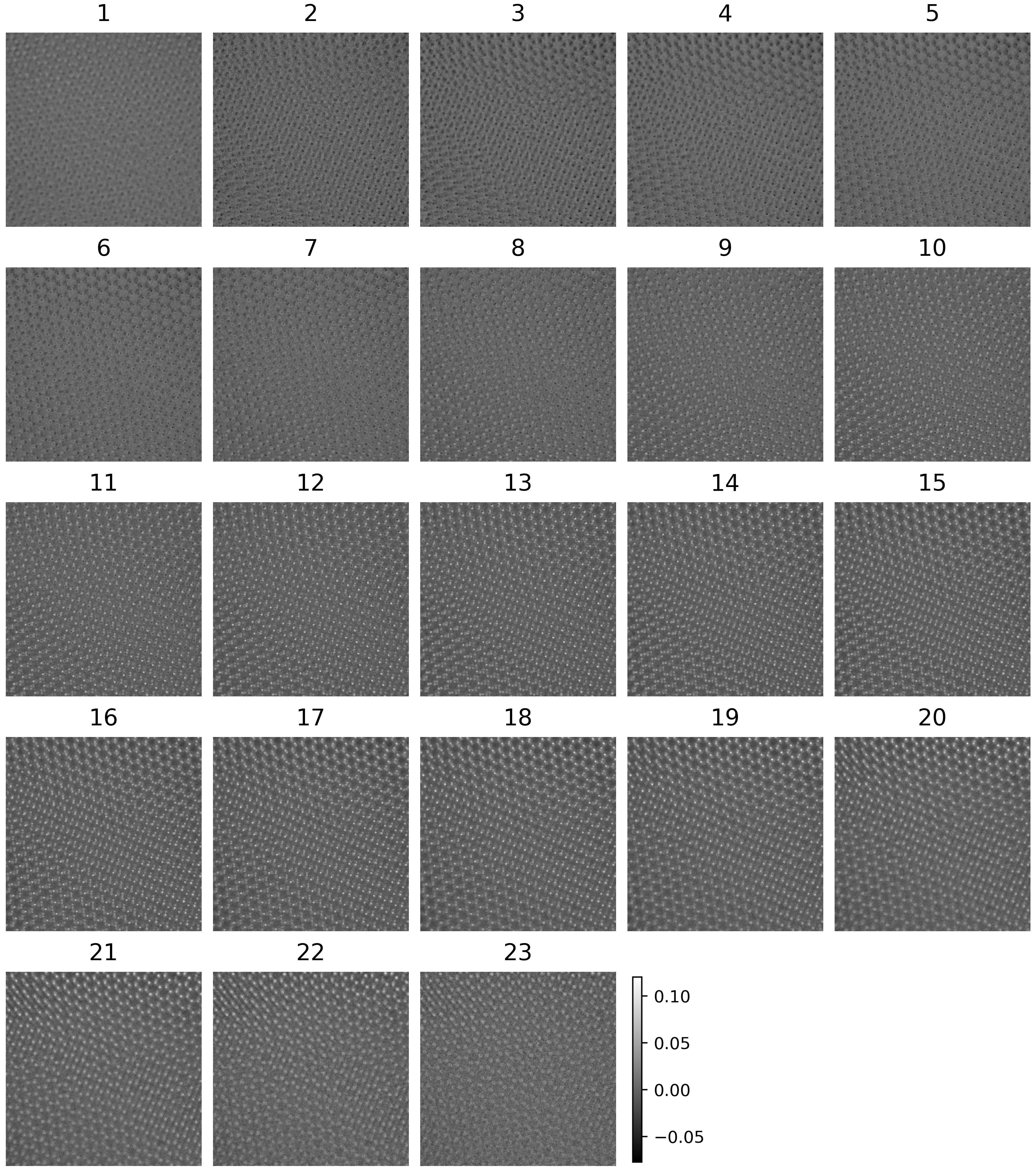}
    \caption{\textbf{Multislice reconstruction with defects.} All slices of the object phase of the ptychographic reconstruction for Fig. 3. The image is 5.52 nm wide and 5.48 nm tall. Phase values are in units of radians.}
    \label{fig:defects_all_slices}
\end{figure}

\begin{figure}[hp!]\centering
    \includegraphics[width=.8\textwidth]{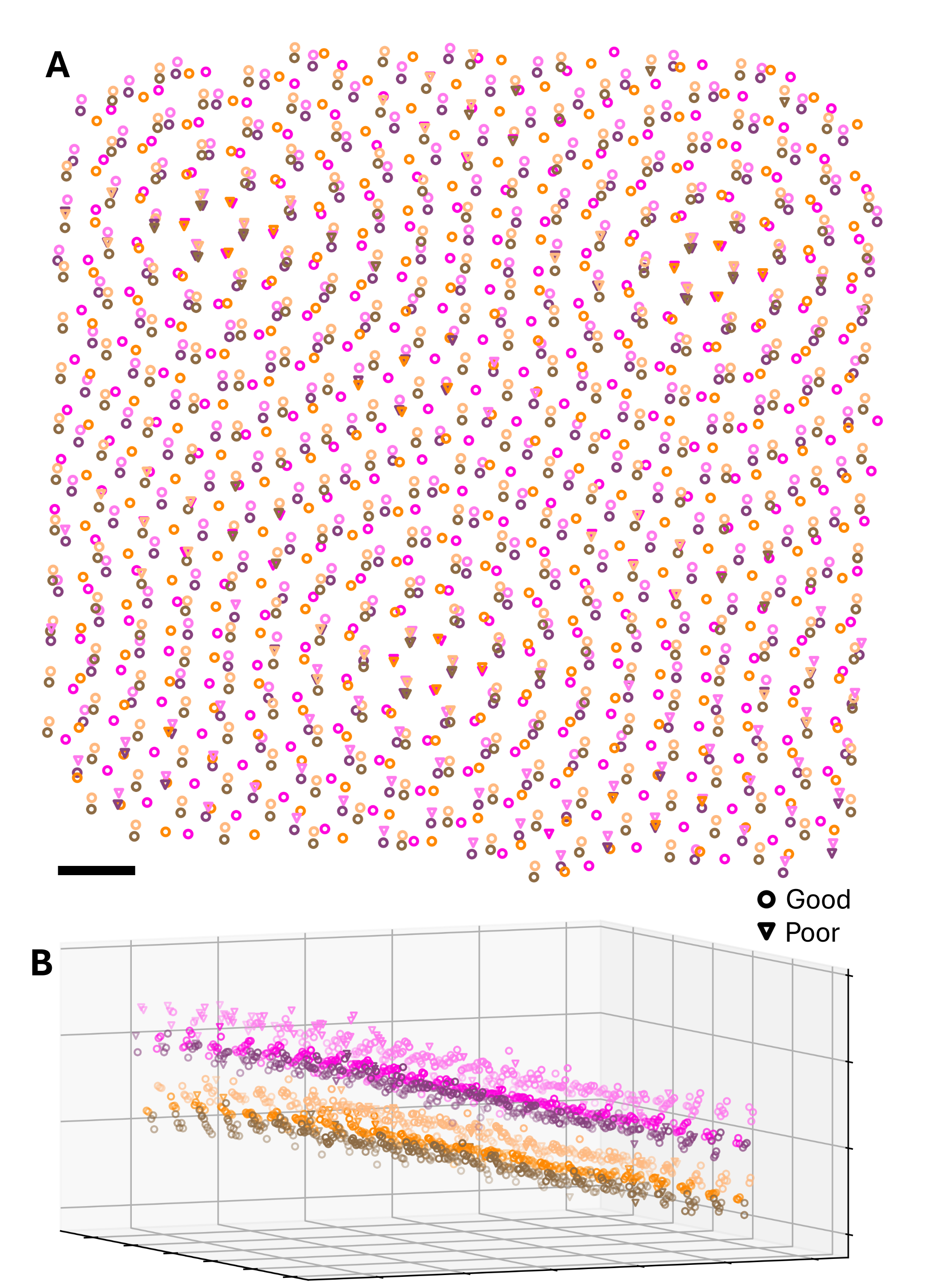}
    \caption{\textbf{Labeling points for coordinate refinement.} Top (\textbf{A}) and oblique (\textbf{B}) view of the raw coordinates extracted from the ptychography simulation (Fig. 2A, fig. S6).  For the sake of the coordinate refinement, the coordinates are labeled as Good (empty circle) or Poor (empty triangle). The coordinate refinement uses a larger weight for the Good coordinates. Scale bar 5~\angstrom. In the oblique view, the grid spacing is 1 nm.}
    \label{fig:ptycho_sim_peak_quality}
\end{figure}

\begin{figure}[hp!]\centering
    \includegraphics[width=.9\textwidth]{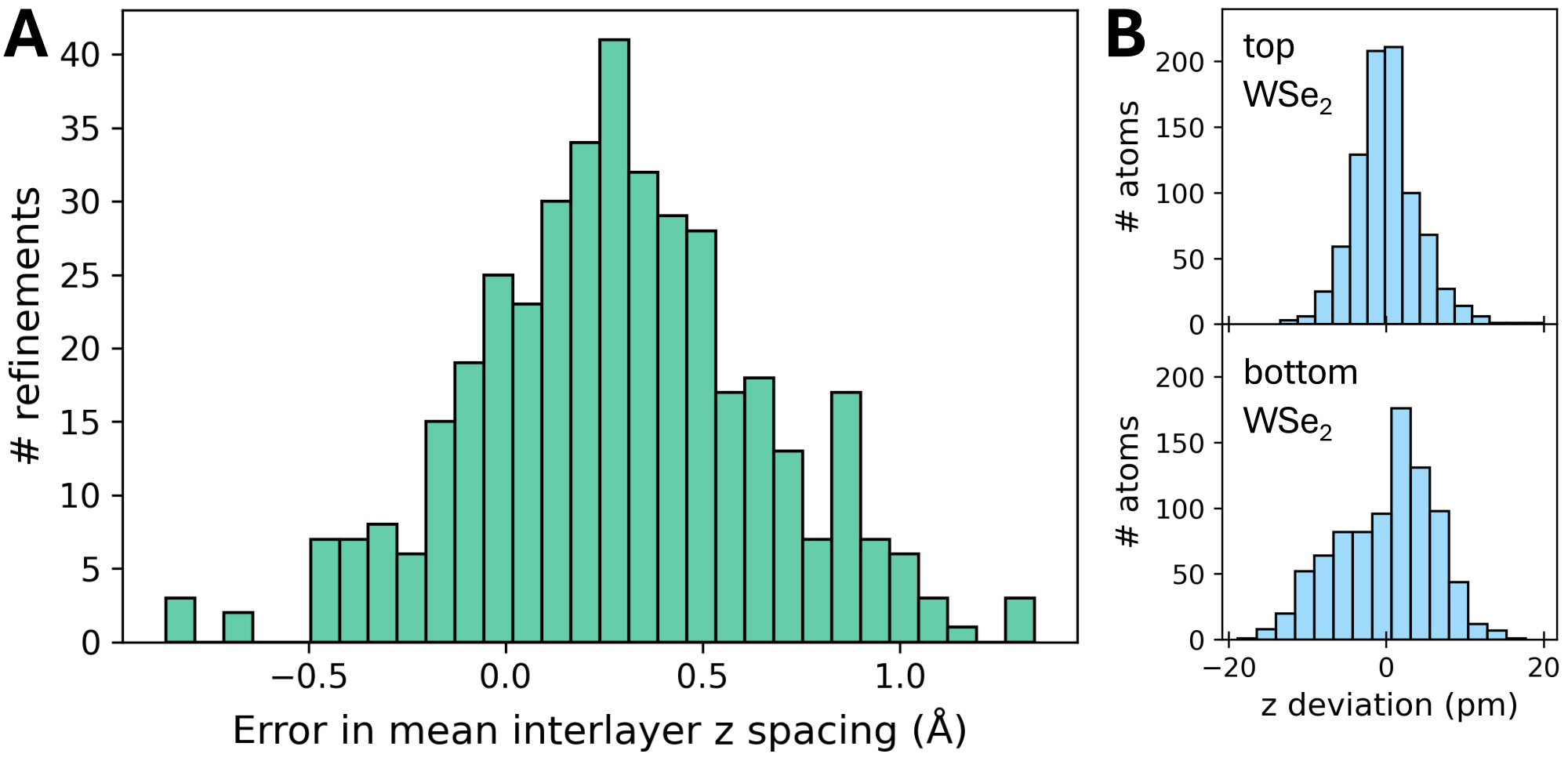}
    \caption{\textbf{$z$ coordinate errors between and within the two \WSe2 layers.} (\textbf{A}) Histogram of the error in the mean interlayer spacing between the two \WSe2 layers along the $z$ direction across 401 separate refinement attempts. (\textbf{B}) Histograms of the deviations of the refined $z$ coordinates of the top and bottom \WSe2 layers, as aligned against the true coordinates.  There are 859 atoms in the top layer and 874 in the bottom.}
    \label{fig:interlayer_z_error_histogram}
\end{figure}

\begin{figure}[hp!]\centering
    \includegraphics[width=.98\textwidth]{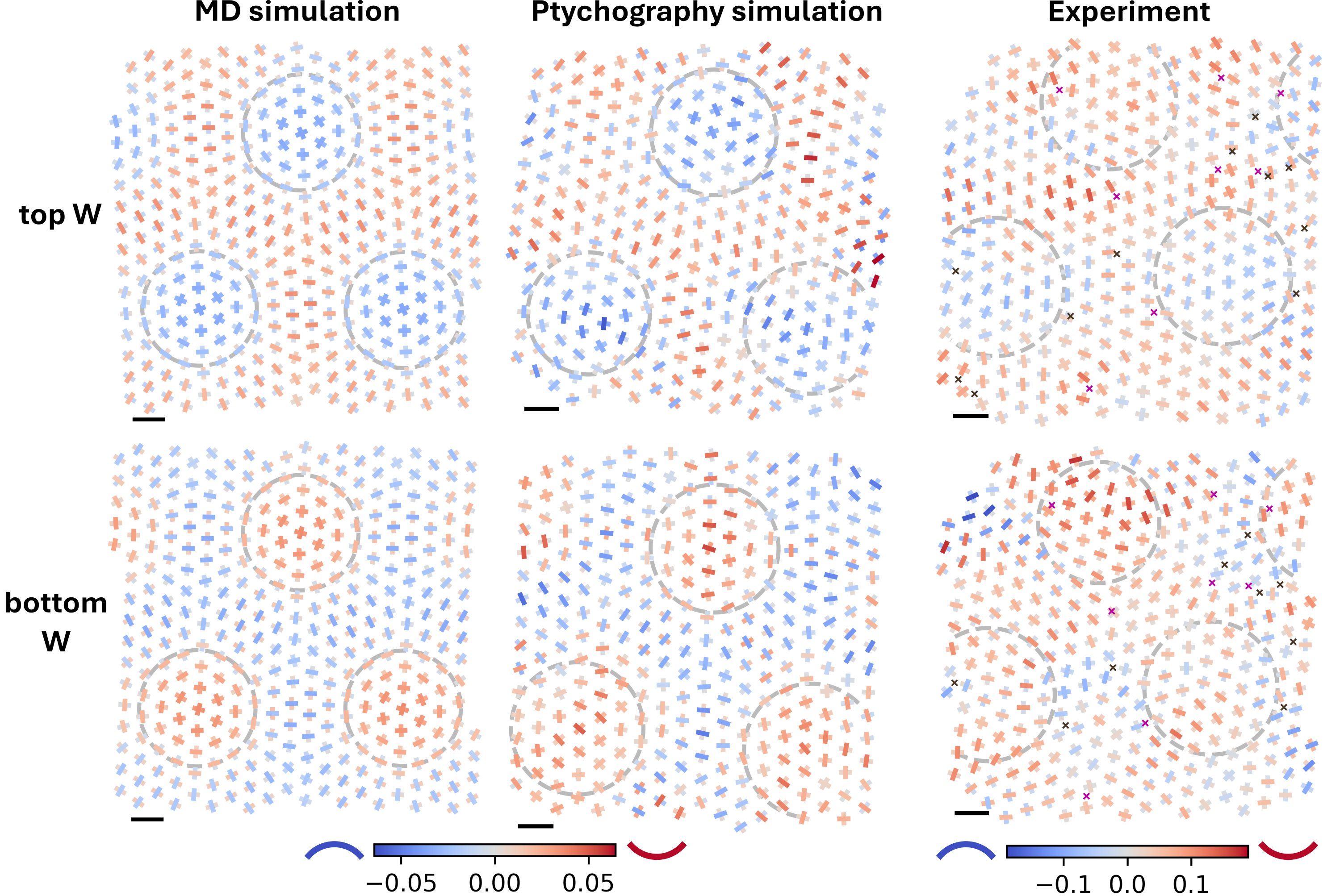}
    \caption{\textbf{Principal curvatures.} Plot of principal curvatures for the top and bottom W layers from three sets of coordinates: MD simulation, ptychography simulation (figs. S6 and S12), and experiment (fig. S1B). The principal curvature direction and value are indicated by thick line segments plotted at each W atom; at each point, there are two principal curvature directions that are orthogonal to each other.  Curvature values are in units of 1/nm. All scale bars are 5~\angstrom. These principal curvature maps are used to calculate the mean curvature maps in Figs. 4C,G and figs. S12C,D.}
    \label{fig:principal_curv}
\end{figure}

\begin{figure}[hp!]\centering
    \includegraphics[width=.95\textwidth]{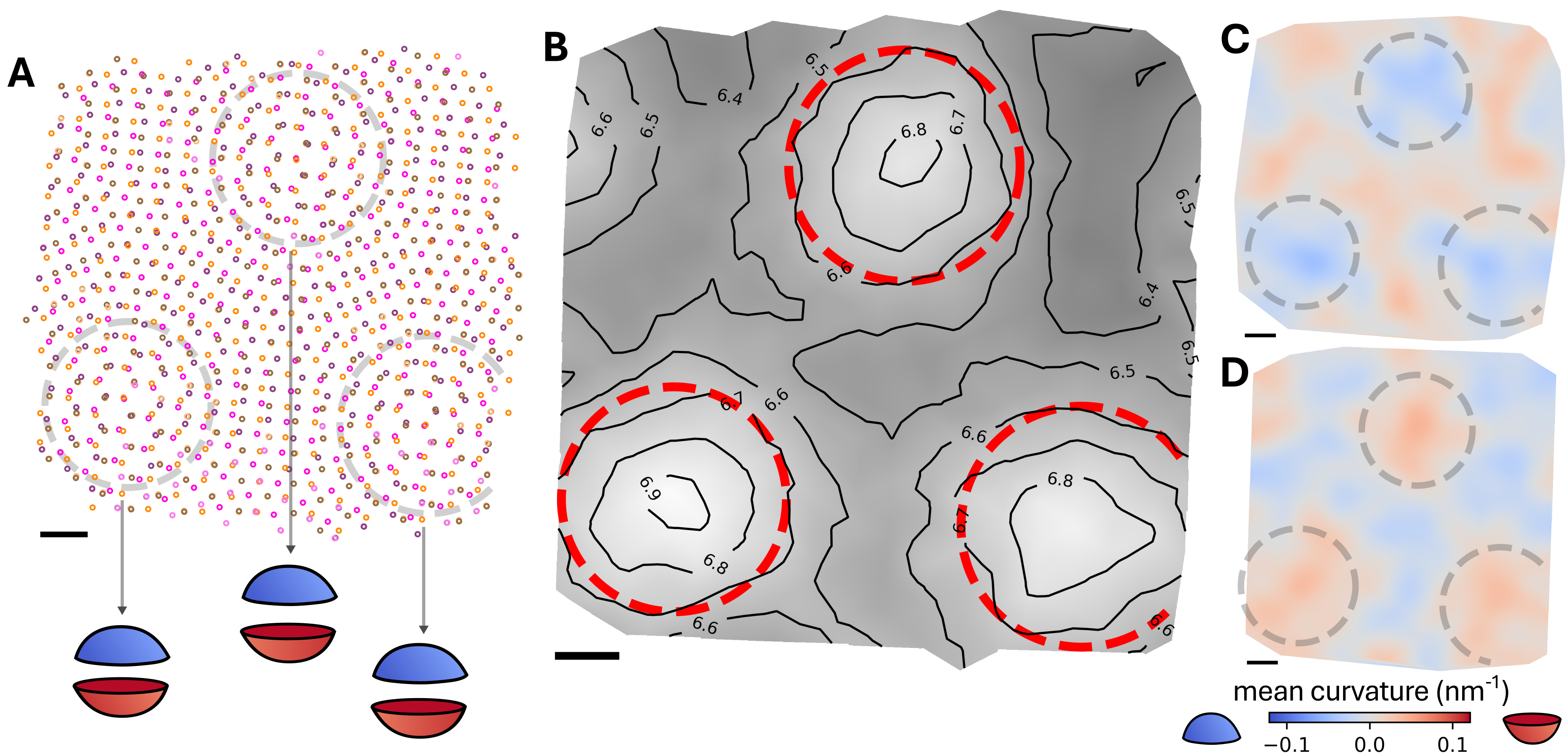}
    \caption{\textbf{Analysis of coordinates from ptychography simulation.} (\textbf{A}) 2D projected coordinates of twisted bilayer \WSe2 obtained from ptychography on simulated 4D-STEM data (fig. S6) using coordinates from MD simulations. All atoms are projected to a common surface defined by the top W atomic layer, to retrieve the stacking order. The locations with approximately AA stacking are labeled with dashed circles. (\textbf{B}) Contour plots of interlayer spacing between the two W layers in angstroms. (\textbf{C} and \textbf{D}) Mean curvature of top (C) and bottom (D) W atomic layers. The curvature of each AA region is also illustrated below (A). Scale bar = 5~\angstrom\ for all panels.}
    \label{fig:ptycho_sim_ils_curv}
\end{figure}




\end{document}